\documentclass[aps,pra,reprint,groupedaddress,superscriptaddress]{revtex4-2}

\usepackage{amsmath}
\usepackage{amssymb}
\usepackage{graphicx}
\usepackage{dcolumn}
\usepackage{bm}
\usepackage[colorlinks,linkcolor=blue,urlcolor=blue,anchorcolor=blue,citecolor=blue]{hyperref}

\begin{document}

\title{Rate and depth of two-axis twisting by pair exchange under the\\
  self-Kerr constraint}

\author{Qing Yu}
\affiliation{School of Information Science and Engineering, Hunan Institute of Engineering, Xiangtan 411104, China}
\author{Ying Li}
\affiliation{School of Information Science and Engineering, Hunan Institute of Engineering, Xiangtan 411104, China}
\author{Cuilu Zhai}
\affiliation{Department of Maths and Physics, Hunan Institute of Engineering, Xiangtan 411104, China}
\author{Shiqing Tang}
\email{sqtang@hynu.edu.cn}
\affiliation{College of Physics and Electronic Engineering, Hengyang Normal University, Hengyang 421002, China}
\author{Wangjun Lu}
\email{wjlu1227@zju.edu.cn}
\affiliation{School of Information Science and Engineering, Hunan Institute of Engineering, Xiangtan 411104, China}
\date{\today}

\begin{abstract}
In this paper, we study two bosonic modes that exchange pairs of quanta,
which realize two-axis twisting exactly: within a sector of fixed particle
number the exchange operator is proportional to the two-axis twisting
generator, with a fixed factor of two.  Because any real nonlinear medium
supplies a local Kerr nonlinearity alongside this operator, the two merge
into one twisting Hamiltonian governed by a single dimensionless ratio
$r=\chi_s/\chi_e$ of the Kerr strength to the exchange strength.  Two
conditions then decide how much of the ideal behavior survives this
constraint, and they must be met separately.  The first concerns speed: the exponential squeezing
rate follows in closed form, vanishes exactly at $r=\pm1$, and recovers as
the square root of the distance from that point.  The second concerns depth:
the smallest attainable squeezing parameter depends only on the anisotropy
of the mean-field saddle, attaining the full two-axis-twisting value at
$r=0$ and $r=\pm3$ regardless of the rate, so that settings sharing the same
anisotropy but differing twofold in rate reach the same depth.  The two
conditions address different targets, and a system may squeeze quickly yet
terminate at a shallow optimum; only $r=0$ is optimal in both respects.  The
closed-form Gaussian solution reveals a further systematic bias: because the
early-time decay rate of the variance exceeds twice the exponential rate, a
fit to short-time data overestimates the rate by more than a factor of two.
Since an isotropic cubic nonlinearity imposes $r=1$, the value at which the
rate vanishes, the two strengths must be mismatched, and a moderate mismatch
already restores most of the optimal depth.  Finally, the constraint is
invisible at low excitation: at two quanta the Kerr term shifts both coupled
states equally, so the two Hamiltonians are indistinguishable until four
quanta.
\end{abstract}

\maketitle

\section{\label{sec:intro}Introduction}

Spin squeezing, which reduces the projection noise of one spin quadrature at
the expense of its conjugate, underlies a large fraction of the proposed
improvements in atomic clocks, magnetometers and matter-wave
interferometers~\cite{Kitagawa1993,Wineland1992,Ma2011,Pezze2018,Toth2014}.
Two generators are standard~\cite{Kitagawa1993}: one-axis twisting (OAT),
which uses the square of a single collective-spin component, and two-axis
twisting (TAT), which uses the difference of the squares of two orthogonal
components.  Their metrological payoffs differ in kind, since OAT improves
the gain only as a fractional power of the particle number whereas TAT
attains the Heisenberg scaling.  The origin of this difference is visible
in the mean-field flow: OAT redistributes phase about the mean spin, whereas
TAT moves the population imbalance by a pair of quanta and thereby breaks the
axial symmetry.

Experimentally, spin squeezing has been demonstrated in internal-state and atom-chip
condensates~\cite{Esteve2008,Gross2010,Riedel2010} and in twin matter
waves~\cite{Lucke2011}, and it now underpins cavity-based quantum
nondemolition measurement~\cite{Leroux2010,SchleierSmith2010,Cox2016},
optically pumped magnetometry~\cite{Sewell2012,Troullinou2021},
squeezed-light-enhanced microwave clocks~\cite{Kruse2016}, and ensembles
addressed on an optical clock
transition~\cite{Pedrozo2020,Robinson2024,Muessel2014,Hosten2016,Greve2022}.
The advantage of two-axis twisting has nevertheless proved difficult to
harvest, and a general twisting tensor was realized only recently by
cavity-mediated engineering~\cite{Luo2025}.  Earlier strategies, including
quenches~\cite{Liu2011}, twist-and-turn dynamics~\cite{Muessel2015},
interaction-based readout~\cite{Strobel2014,Linnemann2016}, Trotterized pulse
sequences and dissipative engineering, achieve two-axis-twisting-like scaling
at the price of additional control or additional decoherence, and
phonon-mediated proposals~\cite{Qiao2026} and studies of twisting under
noise~\cite{Ho2026} face comparable trade-offs.

A more direct route is available in principle: for two bosonic modes at fixed
total occupation, the operator that transfers a pair of quanta from one mode
to the other is the two-axis twisting generator itself, so that neither
Trotterization nor an auxiliary drive nor any term with reversed sign is
required.  The identity is not new, since it underlies exactly solvable
models of two cavities exchanging photon
pairs~\cite{Alexanian2011,Alexanian2011b}, and Hardal and
M\"ustecapl{\i}o{\u g}lu identified it explicitly as the two-axis-twisting
character inherent in two-photon exchange~\cite{Hardal2014}.  The same
identity recurs in the extended Bose--Hubbard literature as pair tunnelling,
observed directly in a double well~\cite{Folling2007} and studied for its
ground-state phases~\cite{Zhu2015}.  In ultracold spinor gases the same
structure is produced by spin-changing collisions, which generate
spin-nematic squeezed vacuum~\cite{Hamley2012}, can be echoed for enhanced
sensing~\cite{Mao2023}, and through the same contact term set the
coherence-time ceiling of spin-exchange collisions~\cite{Huang2023}.
These works either operate at fixed, symmetry-determined couplings or treat
the accompanying mean-field shift as a nuisance; none of them treats that
shift as the second axis of a twisting tensor.

The geometric framework used below is likewise established, a symmetric
twisting tensor being sufficient to characterize any quadratic spin
interaction: the early squeezing rate is set by the total spread of the
tensor eigenvalues, whereas the rate at later times is set by a generally
smaller combination of the eigenvalue spacings.  The two agree exactly when the middle
eigenvalue lies halfway between the extreme ones, so that the dynamics
remains two-axis-twisting-like throughout~\cite{Opatrny2015}.  The
same mean-field-plus-Gaussian machinery has identified two-axis twisting
dynamics in dipolar spinor condensates~\cite{Kajtoch2016}, and the same
geometry pervades anisotropic collective-spin models~\cite{Lipkin1965}.
None of these works, however, addressed the constraint that a Kerr medium
imposes, because the pair-exchange term never appears alone: the local Kerr
nonlinearity is always present with comparable strength, so that, once
multiples of the total spin squared are discarded, the twisting tensor is
locked to a single dimensionless ratio of the Kerr strength to the exchange
strength.

We solve this constrained problem in closed form and obtain three results.
(a)~A rate map: linearizing the exact equations of motion about a coherent
spin state polarized along the intermediate eigenaxis yields an exponential
rate that is a closed-form function of the ratio alone.  This rate vanishes
when the two nonlinearities are equal in magnitude and nowhere else, and it
recovers with a square-root dependence as the ratio moves away from that
value.  (b)~A depth map: the attainable squeezing is a single monotone
function of the anisotropy of the mean-field saddle.  It reaches the full
two-axis-twisting depth at a balanced saddle, where the two transverse
curvatures are equal, and it degrades continuously toward the
one-axis-twisting limit.  (c)~The two maps are independent, the rate
controlling how fast the optimum is reached and the anisotropy controlling
how deep it can become.  Parameter sets that share the same
anisotropy but differ in rate by a sizeable factor yield identical depths at
more than one system size; a realization may therefore possess a high rate
and yet be confined to a shallow optimum.

The same solution yields the transverse covariance in closed form within the
Gaussian description, and it brings one caution: the initial decay rate of
the variance equals twice the exponential rate only for a balanced saddle, so
that a rate extracted from short-time data is overestimated by a large
factor.  We verify this prediction numerically, to within a fraction of a
percent, for every ratio tested.

The corollary for experiment is direct: because an isotropic cubic
nonlinearity makes the Kerr strength equal to the exchange strength, the very
ratio at which the rate vanishes, an experiment must move the ratio away from
unity.  Such control is realistic.  The Kerr
nonlinearity is the resource on which the Kerr-medium route to number-phase
squeezing was founded~\cite{Kitagawa1986} and is routinely engineered and
measured at the single-photon level in superconducting
circuits~\cite{Kirchmair2013,Grimm2020,Lescanne2020}, whereas the combination
required here, a strong multi-photon process together with a tunable Kerr
term, is precisely what nonlinear asymmetric inductive elements provide.
Flux and pump offer in situ access to a dressed operating
point~\cite{Sivak2019,Frattini2018}, and measurements on flux-modulated
nonlinear oscillators show the self- and cross-Kerr terms traded against each
other~\cite{Koenig2026}.  The multi-photon processes
realized so far in these platforms nevertheless convert two photons of one
mode into one photon of another~\cite{Leghtas2015} or transfer pairs between
the two ladders~\cite{Nguyen2024}.  How far the ratio must be moved, and
which platforms can supply the required tuning, is taken up in
Sec.~\ref{sec:disc}.

Finally, because the squeezing parameter compares only second moments, it
cannot distinguish a Gaussian uncertainty ellipse from a bent, strongly
non-Gaussian distribution that shares the same variance, and it is the latter
that limits how long one may profitably twist.  We therefore complement the
variance-based analysis with the covariant spin Wigner
function~\cite{Dowling1994,Brif1999} and the negative-volume
fraction~\cite{Davis2021}, evaluated for both generators at their optima.  At
a moderate particle number the two generators deliver nearly the same
metrological gain but markedly different phase-space structures, and only
within the Gaussian regime can the two pictures be reconciled.

\section{\label{sec:model}Model and exact identities}

Section~\ref{sec:intro} identified pair exchange as an exact realization of
two-axis twisting, with the accompanying Kerr nonlinearity as the
obstruction.  This section makes that obstruction quantitative: we write down
the most general Hamiltonian compatible with the symmetries of the problem
and reduce it to a one-parameter family.

Consider two bosonic modes, $\hat a$ and~$\hat b$.  Their total occupation
$\hat N=\hat n_a+\hat n_b$ is conserved, where $\hat n_a=\hat a^\dagger\hat a$
and $\hat n_b=\hat b^\dagger\hat b$ are the mode occupations.  We work
throughout in the fixed-$N$ sector, whose dimension is $N+1$.
Introduce the Schwinger representation~\cite{Schwinger1952}
\begin{equation}
\begin{aligned}
\hat J_+&=\hat a^\dagger\hat b,\qquad \hat J_-=\hat b^\dagger\hat a,\\
\hat J_z&=\tfrac12(\hat n_a-\hat n_b),\qquad J=\tfrac N2 ,
\end{aligned}
\end{equation}
The Hermitian components are
$\hat J_x=(\hat J_++\hat J_-)/2$ and $\hat J_y=(\hat J_+-\hat J_-)/(2i)$.
They obey $[\hat J_z,\hat J_\pm]=\pm\hat J_\pm$,
$[\hat J_+,\hat J_-]=2\hat J_z$ and
$\hat{\bm J}^2=J(J+1)\hat{\mathbb I}$.  The last relation is the Casimir of
the symmetric irreducible sector,
$\hat{\bm J}^2=\tfrac{\hat N}{2}(\tfrac{\hat N}{2}+1)$, with $\hat N$
replaced by its eigenvalue~$N$.

\begin{figure}
\includegraphics{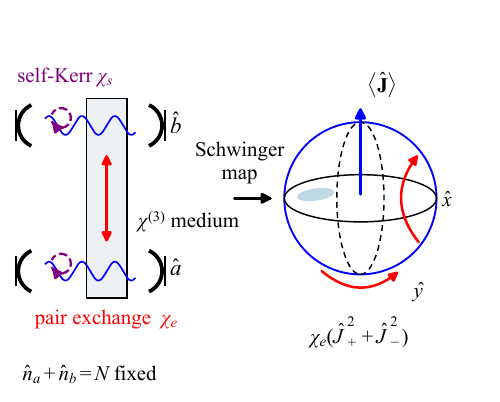}
\caption{\label{fig1}Schematic of the setting.  Two optical
cavities of modes $\hat a$ and $\hat b$ share a cubic nonlinear medium, and
their total occupation $N$ is conserved (left).  Inside that medium the
pair-exchange term $\chi_e$ transfers photon pairs between the two cavities.
Each cavity also acquires a local self-Kerr term $\chi_s$.  Under the
Schwinger map the pair-exchange operator acts on the collective spin as
$\chi_e(\hat J_+^{\,2}+\hat J_-^{\,2})$.  This is the generator of two-axis
twisting about the $\hat x$ and $\hat y$ axes (right).  In the left panel the
pair-exchange term is drawn in red and the self-Kerr term in purple.  It squeezes
the transverse noise ellipse of an initially coherent spin that points along
the intermediate axis.  On the sphere, the solid and dashed
ellipses are the equator and the meridian through the poles (the graticule),
the straight arrow is the mean spin $\langle\hat{\mathbf J}\rangle$, and the
shaded patch is the noise ellipse after twisting.}
\end{figure}

The Hamiltonian is built from these operators; taking the most general form
that contains four field operators and requiring it to conserve $\hat N$, to
be symmetric under $a\leftrightarrow b$ and to contain no linear tunnelling
selects the form
\begin{align}
\hat H={}&U\bigl(\hat a^{\dagger2}\hat a^2+\hat b^{\dagger2}\hat b^2\bigr)
+U_c\,\hat n_a\hat n_b
+\chi_e\bigl(\hat a^{\dagger2}\hat b^2+\hat a^2\hat b^{\dagger2}\bigr) .
\label{eq:Hgen}
\end{align}
Here $U$ is the self-Kerr coefficient of one mode, $U_c$ the cross-Kerr
coefficient and $\chi_e$ the strength of the pair exchange.  Inside the
fixed-$N$ sector, where $\hat N$ may be replaced throughout by its
eigenvalue~$N$, the following three operator identities are exact:
\begin{subequations}
\label{eq:ids}
\begin{align}
\hat S&\equiv \hat a^{\dagger2}\hat a^2+\hat b^{\dagger2}\hat b^2
      = 2\hat J_z^2+\tfrac{\hat N^2}{2}-\hat N , \label{eq:idS}\\
\hat n_a\hat n_b&= \tfrac{\hat N^2}{4}-\hat J_z^2 , \label{eq:idC}\\
\hat E&\equiv \hat a^{\dagger2}\hat b^2+\hat a^2\hat b^{\dagger2}
      = \hat J_+^2+\hat J_-^2=2(\hat J_x^2-\hat J_y^2). \label{eq:idE}
\end{align}
\end{subequations}
Equations~(\ref{eq:idS})--(\ref{eq:idE}) follow by direct substitution, as
does the $\mathrm{su}(2)$ relation $[\hat J_x,\hat J_y]=i\hat J_z$.
The setting and the mapping are summarized in Fig.~\ref{fig1}.

Two consequences of Eqs.~(\ref{eq:ids}) are immediate.  First, any
combination of self-Kerr and cross-Kerr produces a term proportional to
$\hat J_z^2$ and nothing else, so that no stacking of Kerr nonlinearities can
generate TAT; this is the fixed-$N$ version of the elementary statement that
any Hamiltonian diagonal in the Fock basis is a function of $\hat J_z$ and
therefore causes pure dephasing.  Second, self-Kerr and cross-Kerr are not
independent within the sector, which is what makes the one-parameter
description exact rather than merely convenient: combining
Eqs.~(\ref{eq:idS}) and~(\ref{eq:idC}), the whole Kerr
sector of~(\ref{eq:Hgen}) collapses to a single twisting coefficient,
\begin{equation}
\begin{aligned}
\hat H &= \lambda_x\hat J_x^2+\lambda_y\hat J_y^2+\lambda_z\hat J_z^2
+f(\hat N),\\
\bm\lambda&=(2\chi_e,\,-2\chi_e,\,2\chi_s),
\end{aligned}
\label{eq:Hdiag}
\end{equation}
where
\begin{equation}
\chi_s \equiv U-\tfrac12 U_c ,
\label{eq:chis}
\end{equation}
and the function $f(\hat N)$ collects all terms that are constant inside the
sector.  Up to an overall scale the dynamics therefore depends on the single
dimensionless ratio
\begin{equation}
r\equiv \frac{\chi_s}{\chi_e}=\frac{U-\frac12U_c}{\chi_e},
\end{equation}
and on nothing else.  Throughout we fix the total nonlinearity budget,
$\sqrt{\chi_s^2+\chi_e^2}=1$, quote all rates per unit of available
nonlinearity and measure time in units of the inverse nonlinearity, so that
the squeezing may be quantified by the Wineland parameter~\cite{Wineland1992}
\begin{equation}
\xi^2=\frac{N\,\min_{\hat n\perp\langle\hat{\bm J}\rangle}
      \mathrm{Var}(\hat J_{\hat n})}{|\langle\hat{\bm J}\rangle|^2},
\label{eq:xi2def}
\end{equation}
Here $\hat J_{\hat n}\equiv\hat{\bm n}\cdot\hat{\bm J}$ is the collective
spin $\hat{\bm J}=(\hat J_x,\hat J_y,\hat J_z)$ projected on the unit vector
$\hat{\bm n}$, and the variance is minimized over directions transverse to
the mean spin $\langle\hat{\bm J}\rangle$, so that $\xi^2=1$ by construction
for a coherent spin state.  The metrological gain is usually quoted in
decibels (dB) as $-10\log_{10}\xi^2$.  We call the smallest value reached
during the time evolution the attainable depth, denoted $\xi^2_{\min}$, and
the time at which it is reached the optimal time $t_{\rm opt}$.

Two limiting cases are recorded here because they recur throughout.  When
$\chi_e=0$ the ratio diverges and the dynamics reduces to OAT about~$z$,
whereas $r=1$ gives
\begin{equation}
\hat S+\hat E = \hat N^2-4\hat J_y^2 ,
\label{eq:oatnat}
\end{equation}
so that an equal-weight mixture of self-Kerr and pair exchange is pure OAT,
about the circular-polarization axis.  This case is not marginal, since an
isotropic medium produces exactly it: Appendix~\ref{app:chi3} shows that two
modes sharing a spatial profile in a medium with an isotropic third-order
susceptibility $\chi^{(3)}$ give this mixture~\cite{GarciaFerrer2007}, so
that the constraint studied here is binding rather than optional.

The model is thereby reduced to the single-parameter family of
Eq.~(\ref{eq:Hdiag}), parametrized by~$r$.  Section~\ref{sec:mf} determines
how fast a member of that family squeezes an initially coherent spin, by
linearizing the exact equations of motion about that state.

\section{\label{sec:mf}Mean-field linearization and the squeezing rate}

Section~\ref{sec:model} reduced the dynamics to a one-parameter family of
twisting tensors, and the next step is to determine how fast such a tensor
squeezes an initially coherent spin.  We proceed in two stages, first
linearizing the exact equations of motion about the initial state and then
classifying the resulting linear flow, which yields both the exponential rate
and, as a by-product, the correct choice of initial axis.

\subsection{Equations of motion and the choice of initial axis}

For a Hamiltonian $\hat H=\lambda_x\hat J_x^2+\lambda_y\hat J_y^2+\lambda_z
\hat J_z^2$ the Heisenberg equations are exact and bilinear:
\begin{equation}
\frac{d\hat J_x}{dt}=2(\lambda_y-\lambda_z)\,\overline{\hat J_y\hat J_z},
\qquad \text{and cyclically},
\label{eq:eom}
\end{equation}
where $\overline{AB}\equiv\frac12(AB+BA)$.  We prepare a coherent spin
state (CSS) polarized along a principal axis $k$ of the twisting tensor, so
that it is an eigenstate of $\hat J_k$ with eigenvalue~$+J$, and we linearize
the transverse fluctuations by replacing $\hat J_k$ with~$J$.  Which of the
three axes is to be chosen follows from the character of the resulting
flow.  Writing $(i,j,k)$ for
a right-handed cyclic permutation of $(x,y,z)$, we find that the transverse
pair obeys
\begin{equation}
\frac{d}{dt}\begin{pmatrix} \hat J_i\\ \hat J_j\end{pmatrix}
=\begin{pmatrix}0&p\\ q&0\end{pmatrix}
\begin{pmatrix} \hat J_i\\ \hat J_j\end{pmatrix},
\quad
\begin{aligned}
p&=2J(\lambda_j-\lambda_k),\\
q&=2J(\lambda_k-\lambda_i).
\end{aligned}
\label{eq:lin}
\end{equation}
The eigenvalues of this matrix are $\pm\sqrt{pq}$, and three cases arise.  If
$pq<0$ they are purely imaginary and the coherent spin state simply rotates,
the covariance ellipse being transported rigidly with no squeezing at all, as
happens whenever the state is polarized along an extremal eigenaxis.  If
$pq=0$ they are degenerate and the flow is a shear, so that the variance
grows only as a power law; this is the critical, one-axis-twisting-like case.
If $pq>0$ they are real and opposite, $\pm\Lambda$, with
\begin{equation}
\Lambda=\sqrt{pq}
=2J\sqrt{(\lambda_{\max}-\lambda_{\rm mid})(\lambda_{\rm mid}-\lambda_{\min})},
\label{eq:Lambda}
\end{equation}
where $\lambda_{\max}\ge\lambda_{\rm mid}\ge\lambda_{\min}$ are the three
components of $\bm\lambda$ arranged in non-increasing order.  The flow is then
hyperbolic, and because the symplectic eigenvalue of the covariance is
conserved, one transverse quadrature is squeezed at the exponential
rate~$\Lambda$ while its conjugate is anti-squeezed at the same rate.

Since only the third case produces exponential squeezing, the choice of
initial state is settled: the coherent spin state must point along the
intermediate eigenaxis, only $k={\rm mid}$ giving $pq>0$, whereas choosing
either extremal axis produces $pq<0$ and no squeezing at all even though the
Hamiltonian is the same.  Equation~(\ref{eq:Lambda}) is the first main
result of this work, and it is invariant under the addition of any multiple
of $\hat J^2$ to~$\hat H$, which shifts all three components of
$\bm\lambda$ equally and therefore leaves the equation unchanged, as it must.

\subsection{Rate and anisotropy for this family}

\begin{figure}
\includegraphics{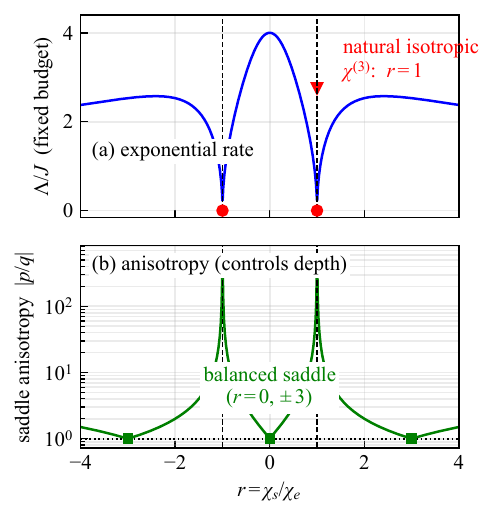}
\caption{\label{fig2}(a) Exponential squeezing rate per unit nonlinearity
budget, Eq.~(\ref{eq:Lamofr}), versus the Kerr ratio $r=\chi_s/\chi_e$.  The
zero at $r=\pm1$ is exact and square-root-like,
$\Lambda\simeq4J\sqrt{|\,1-|r|\,|}$.  Red dots sit on the two zeros and
black dashed verticals mark $r=\pm1$; the red arrow above the upper dot
singles out $r=1$, the value imposed by an isotropic
$\chi^{(3)}$ (Appendix~\ref{app:chi3}).  (b) Saddle
anisotropy $\mathcal A=\max(|p/q|,|q/p|)$ on a logarithmic ordinate.  Its
minimum, $\mathcal A=1$, is attained only at $r\in\{0,\pm3\}$, where it is
marked by green squares and by the dotted horizontal guide at
$\mathcal A=1$; it diverges at the critical points $r=\pm1$, again marked by
black dashed verticals.}
\end{figure}

Specializing Eq.~(\ref{eq:Lambda}) to the family of
Sec.~\ref{sec:model} by inserting $\bm\lambda=(2\chi_e,-2\chi_e,2\chi_s)$ and
removing the budget normalization, we find that the rate takes a closed form
in $|r|=\rho$,
\begin{equation}
\frac{\Lambda}{J}=
\begin{cases}
\dfrac{4\sqrt{1-\rho^2}}{\sqrt{1+\rho^2}}, & \rho\le 1,\\[2ex]
\dfrac{4\sqrt{2(\rho-1)}}{\sqrt{1+\rho^2}}, & \rho\ge 1,
\end{cases}
\label{eq:Lamofr}
\end{equation}
displayed in Fig.~\ref{fig2}(a).  The rate is even in~$r$, as the $a\leftrightarrow b$
symmetry requires, since exchanging $\hat a$ and $\hat b$ maps $r$ to~$-r$
and leaves every second moment unchanged up to a rotation.

Equation~(\ref{eq:Lamofr}) vanishes exactly at $r=\pm1$, and expanding
about $\rho=1$ gives
\begin{equation}
\Lambda\simeq 4J\,\sqrt{|\,1-\rho\,|}\qquad (\rho\to1),
\end{equation}
so that the rate recovers as a square root: a small departure from the
critical ratio already restores a substantial fraction of the optimal rate.
This vanishing has a simple operator origin: at $r=\pm1$ the twisting tensor
becomes $\bm\lambda\propto(2,-2,\pm2)$ and, because $\hat J^2$ is constant
inside the sector, the Hamiltonian reduces to $\hat H\propto-4\hat J_y^2$ at
$r=+1$ and to $\hat H\propto+4\hat J_x^2$ at $r=-1$, so that the dynamics is
rigorously one-axis twisting about the~$y$ axis and the~$x$ axis
respectively.  This is the operator counterpart of Eq.~(\ref{eq:oatnat}), and
it explains why $\xi^2$ then improves only as~$N^{-2/3}$.

The saddle anisotropy follows from Eq.~(\ref{eq:lin}):
\begin{equation}
\mathcal A\equiv\max\!\left(\left|\frac pq\right|,\left|\frac qp\right|\right)
=
\begin{cases}
\dfrac{1+\rho}{1-\rho}, & \rho<1,\\[2ex]
\max\!\left(\dfrac{\rho-1}{2},\ \dfrac{2}{\rho-1}\right), & \rho>1,
\end{cases}
\label{eq:aniso}
\end{equation}
plotted in Fig.~\ref{fig2}(b).  The rate therefore singles out
$r=\pm1$ as the forbidden values, whereas the anisotropy singles out
$r\in\{0,\pm3\}$ as the optimal values.  Which of the two governs the
attainable depth is determined in Sec.~\ref{sec:depth}: there the anisotropy,
and not the rate, sets the ceiling.  Before that, we exploit the linearity of
Eq.~(\ref{eq:lin}) to solve the whole transient in closed form.

\section{\label{sec:gauss}Closed-form Gaussian dynamics}

Section~\ref{sec:mf} gave the asymptotic rate but not the transient that
leads to it.  Because the linearized equations are exactly solvable, that
transient is available analytically, and it serves two purposes: it locates
the optimal time and depth within the Gaussian approximation, and it
identifies a systematic error in the extraction of rates from early-time
data.

\subsection{Exact solution of the transverse covariance}

Because Eq.~(\ref{eq:lin}) is linear, the transverse covariance matrix can be
obtained in closed form; with $C_0=\frac J2\,\mathbb I$ for the initial CSS,
\begin{equation}
C(t)=e^{Mt}\,C_0\,e^{M^{\mathsf T}t},\qquad
M=\begin{pmatrix}0&p\\ q&0\end{pmatrix}.
\end{equation}
Two invariants follow immediately from $\operatorname{tr}M=0$, and together
they fix both eigenvalues:
\begin{equation}
\det C(t)=\left(\frac J2\right)^2\quad\text{(conserved)},\qquad
\operatorname{tr}C(t)=J\,T(t),
\end{equation}
with
\begin{equation}
\begin{aligned}
T(t)&=1+(1+\alpha)\,\sinh^2(\Lambda t),\\
\alpha&\equiv\frac12\!\left(\frac pq+\frac qp\right)
=\frac12\!\left(\mathcal A+\frac1{\mathcal A}\right)\ge1 .
\end{aligned}
\label{eq:Tdef}
\end{equation}
Solving for the eigenvalues,
\begin{equation}
\begin{aligned}
v_{\min,\max}&=\frac J2\left(T\mp\sqrt{T^2-1}\right),\\
v_{\min}&=\frac{J/2}{T+\sqrt{T^2-1}} .
\end{aligned}
\label{eq:vmin}
\end{equation}
The Wineland squeezing parameter follows from
$\xi^2=N\,v_{\min}/|\langle\hat{\bm J}\rangle|^2$; using
$\langle\hat{\bm J}^2\rangle=J(J+1)$ and neglecting the (small) variance along
the mean-spin direction gives
\begin{equation}
|\langle\hat{\bm J}\rangle|^2=J(J+1)-J\,T(t),
\end{equation}
so that, finally,
\begin{equation}
\xi^2(t)=\frac{N\,\dfrac J2\left[T-\sqrt{T^2-1}\right]}
              {J(J+1)-J\,T(t)},\qquad
T(t)\ \text{from Eq.}~(\ref{eq:Tdef}).
\label{eq:xi2closed}
\end{equation}
This is the central analytic result of this section, expressing the whole
transient in terms of two parameters, the rate~$\Lambda$ and the anisotropy
parameter~$\alpha$.  As a first check, at $t=0$ one has $T=1$ and
$\xi^2=1$ exactly, as required.

\subsection{Two limits, and why the short-time slope is misleading}

The solution contains two distinct decay rates, and confusing them is the
main practical hazard: expanding Eq.~(\ref{eq:vmin}) for small $t$ gives
$T-1=(1+\alpha)\Lambda^2t^2$ and hence
\begin{equation}
\left.\frac{d\ln v_{\min}}{dt}\right|_{t\to0}
=-\Lambda\sqrt{2(1+\alpha)}=-\,|p+q| ,
\label{eq:slope0}
\end{equation}
where the last equality uses $|p+q|=\Lambda\,|\sqrt{p/q}+\sqrt{q/p}|$.  In the
opposite limit, $T\gg1$ implies $v_{\min}\simeq J/(4T)\propto e^{-2\Lambda t}$
and
\begin{equation}
\frac{d\ln v_{\min}}{dt}\xrightarrow[t\to\infty]{}-2\Lambda .
\label{eq:slopeinf}
\end{equation}
By the arithmetic--geometric mean inequality,
$|p+q|\ge2\sqrt{pq}=2\Lambda$, with equality only when $p=q$; the initial
decay rate of the variance is thus $|p+q|$, which coincides with
$2\Lambda$ only for a balanced saddle, that is, at $r\in\{0,\pm3\}$.  A
slope fitted at short times therefore overestimates $\Lambda$ by the factor
$\sqrt{2(1+\alpha)}/2$, which already reaches $2.29$ at $r=0.9$.

\begin{figure}
\includegraphics{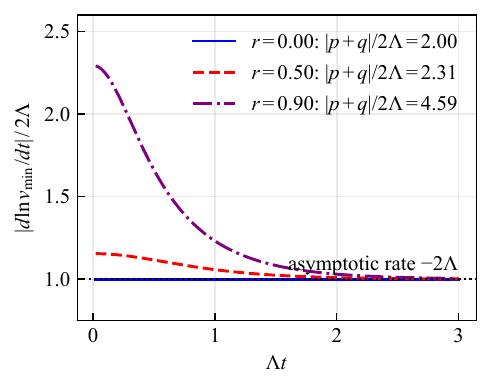}
\caption{\label{fig3}Crossover of the instantaneous decay rate from its
short-time value $|p+q|$ to its asymptotic value $2\Lambda$.  Curves are the
closed-form prediction of Eq.~(\ref{eq:vmin}); open symbols are exact
diagonalization at $N=800$.  Only at $\mathcal A=1$ does the curve start at
$1$; otherwise the early dynamics is faster than the true exponential rate,
which is the origin of the systematic overestimate discussed in the text.
The dotted horizontal line is the asymptote, reached by every curve once
$\Lambda t\gg1$.}
\end{figure}

Figure~\ref{fig3} shows this crossover for three values of~$r$, the analytic
curves and the exact numerical data agreeing over the whole window before
saturation.  A practical rule follows: the decay should be fitted where
$T\gtrsim5$, equivalently $\Lambda t\gtrsim1$, because the dependence on
$\alpha$ has become negligible there, or alternatively the full
Eq.~(\ref{eq:xi2closed}) may be fitted with $\alpha$ fixed by
Eq.~(\ref{eq:aniso}).  A second caution concerns the fitting window: past the
optimal time the squeezing parameter rises again and later falls steeply, and
because that later fall is unrelated to~$\Lambda$, any sliding-window fit
must stop at~$t_{\rm opt}$.

\subsection{Optimal time and optimal depth for the balanced saddle}

Having established the transient, we locate its minimum, which follows from
$d[(J-T)(T+\sqrt{T^2-1})]/dT=0$, i.e.\ $\sqrt{T^2-1}=J-T$, whence
\begin{equation}
T_{\rm opt}=\frac{J^2+1}{2J}\simeq \frac J2 .
\end{equation}
For $\mathcal A=1$ ($\alpha=1$) this gives the two classic TAT results in
closed form,
\begin{equation}
\xi^2_{\min}=\frac{4}{N}\quad(\mathcal A=1),\qquad
\Lambda t_{\rm opt}\simeq\frac12\ln\!\frac{2J}{1+\alpha}
\xrightarrow{\alpha=1}\frac12\ln J .
\label{eq:optTAT}
\end{equation}
Thus the optimal time scales as $\ln J/\Lambda$ rather than as a power of~$N$,
the hallmark of exponential rather than algebraic squeezing.

\begin{table}[b]
\caption{\label{tab:model}Accuracy of the closed-form solution,
Eqs.~(\ref{eq:vmin})--(\ref{eq:xi2closed}), for $r=0$, $N=800$.  The Gaussian
description is excellent while $T\lesssim10$ and fails, as expected, once the
fluctuation spread becomes comparable to $J$ itself.}
\begin{ruledtabular}
\begin{tabular}{ c c d d c }
\multicolumn{1}{c}{$t$}&
\multicolumn{1}{c}{$T$}&
\multicolumn{1}{c}{$v_{\min}$ (exact)}&
\multicolumn{1}{c}{$v_{\min}$ (analytic)}&
\multicolumn{1}{c}{ratio}\\
\colrule
$2.0\times10^{-4}$ &    1.21 & 105.527 & 105.458 & 1.001\\
$5.0\times10^{-4}$ &    2.58 &  40.459 &  40.379 & 1.002\\
$1.0\times10^{-3}$ &   12.29 &   8.257 &   8.152 & 1.013\\
$1.5\times10^{-3}$ &   60.76 &   1.779 &   1.646 & 1.081\\
$1.99\times10^{-3}$&  287.74 &   0.548 &   0.348 & 1.577\\
\end{tabular}
\end{ruledtabular}
\end{table}

Table~\ref{tab:model} quantifies the accuracy, and the breakdown at
$T\to J$ is not a defect of the calculation but a physical statement: at the
optimal time the anti-squeezed quadrature has already spread to
$v_{\max}\simeq J^2/2$, so that the state is strongly non-Gaussian and is
appreciably wrapped around the Bloch sphere.  For the same reason the depth
at large anisotropy lies beyond the reach of Eq.~(\ref{eq:xi2closed}) and
requires the exact treatment of Sec.~\ref{sec:depth}.

\section{\label{sec:depth}Attainable depth: exact numerical results}

Because the Gaussian solution of Sec.~\ref{sec:gauss} is quantitative only
while $T\lesssim10$, as Table~\ref{tab:model} documents, and because the
attainable depth is reached precisely when the fluctuations have grown to
that scale, the depth itself has to be obtained from the exact dynamics.
This section does so by direct diagonalization and then tests the outcome
against the prediction of Sec.~\ref{sec:mf} that the depth is a function of
the anisotropy alone.

\subsection{Method}

We work in the fixed-$N$ sector of dimension $N+1$, constructing
$\hat S$ and $\hat E$ directly in the Fock basis, forming
$\hat H=\chi_s\hat S+\chi_e\hat E$ and diagonalizing once.  The CSS is the
$+J$ eigenstate of $\hat J_k$, with $k$ the intermediate eigenaxis of
$\bm\lambda$ as dictated by Eq.~(\ref{eq:Lambda}); we verified that using an
extremal axis indeed yields no squeezing.  The time evolution $\exp(-i\hat Ht)$
is evaluated in the energy eigenbasis, and at each time we
compute the full $3\times3$ symmetrized covariance of $\hat{\bm J}$, project
it onto the plane transverse to $\langle\hat{\bm J}\rangle$ and evaluate
Eq.~(\ref{eq:xi2def}).  The minimum over $t$ is located by a logarithmic scan followed by local
refinement.  Results are reported for $N=50,100,200,400,800$, the largest
matrix diagonalized being $801\times801$; further details of the numerical
procedure are given in Appendix~\ref{app:num}.

\subsection{$r=0$ and $r=\pm3$ are unitarily equivalent}

We first test the prediction that the depth is governed by the anisotropy.
Since the anisotropy is minimal, $\mathcal A=1$, at $r\in\{0,\pm3\}$, these
three points must share a single depth even though their rates differ, and the
following argument shows that they share the same time evolution.

\begin{table}[t]
\caption{\label{tab:rate}Rates: theory versus exact numerics at $N=800$
($J=400$).  $\Lambda_{\rm th}$ is Eq.~(\ref{eq:Lamofr}) with the budget
$\sqrt{\chi_s^2+\chi_e^2}=1$; $\Lambda_{\rm asym}$ is extracted from the slope
of $\ln v_{\min}$ in the asymptotic window $5\le T\le50$.  The intermediate
eigenaxis, which is the required CSS polarization, is $z$ for $|r|<1$, $x$ for
$r>1$ and $y$ for $r<-1$ (degenerate at $|r|=1$); all entries are even in $r$.}
\begin{ruledtabular}
\begin{tabular}{ c c c c c c }
\multicolumn{1}{c}{$r$}&
\multicolumn{1}{c}{$\mathcal A$}&
\multicolumn{1}{c}{$\alpha$}&
\multicolumn{1}{c}{$\Lambda_{\rm th}$}&
\multicolumn{1}{c}{$\Lambda_{\rm asym}$}&
\multicolumn{1}{c}{ratio}\\
\colrule
 0    &   1     & 1      & 1600.0 & 1563.7 & 0.977\\
 0.25 &   1.667 & 1.133  & 1502.9 & 1474.9 & 0.981\\
 0.5  &   3     & 1.667  & 1239.4 & 1235.9 & 0.997\\
 0.75 &   7     & 3.571  &  846.6 &  888.6 & 1.050\\
 0.9  &  19     & 9.526  &  518.4 &  615.3 & 1.187\\
 1.0  &$\infty$ & ---    &    0   &  ---   & ---   \\
 1.1  &  20     & 10.025 &  481.3 &  576.2 & 1.197\\
 1.5  &   4     & 2.125  &  887.5 &  896.7 & 1.010\\
 3.0  &   1     & 1      & 1011.9 &  988.9 & 0.977\\
\end{tabular}
\end{ruledtabular}
\end{table}

\begin{table}[b]
\caption{\label{tab:depth}Short-time slope versus the prediction
Eq.~(\ref{eq:slope0}), and attainable depth at $N=800$, the latter quoted in
dB, that is, as $-10\log_{10}\xi^2_{\min}$.  The last column is
the fitted exponent of $\xi^2_{\min}\propto N^{-p}$ over $N=50$--$800$.}
\begin{ruledtabular}
\begin{tabular}{ c c c c c c }
\multicolumn{1}{c}{$r$}&
\multicolumn{1}{c}{$2\Lambda$}&
\multicolumn{1}{c}{$|p+q|$}&
\multicolumn{1}{c}{slope}&
\multicolumn{1}{c}{$\frac{\rm slope}{|p+q|}$}&
\multicolumn{1}{c}{dB / $p$}\\
\colrule
 0    & 3200.0 & 3200.0 & $-3196.0$ & 0.999 & 23.13 / 0.974\\
 0.25 & 3005.9 & 3104.5 & $-3100.6$ & 0.999 & 23.03 / 0.972\\
 0.5  & 2478.7 & 2862.2 & $-2858.6$ & 0.999 & 22.68 / 0.964\\
 0.75 & 1693.3 & 2560.0 & $-2556.8$ & 0.999 & 21.88 / 0.935\\
 0.9  & 1036.8 & 2378.5 & $-2375.6$ & 0.999 & 20.73 / 0.874\\
 1.1  &  962.7 & 2260.2 & $-2257.4$ & 0.999 & 20.67 / 0.871\\
 1.5  & 1775.0 & 2218.8 & $-2216.0$ & 0.999 & 22.44 / 0.957\\
 3.0  & 2023.9 & 2023.9 & $-2021.3$ & 0.999 & 23.13 / 0.974\\
\colrule
 1.0  &    0   &  ---   & algebraic & ---   & 18.54 / 0.724\\
\end{tabular}
\end{ruledtabular}
\end{table}

Tables~\ref{tab:rate} and~\ref{tab:depth} collect the principal results, and
they support two conclusions.  First, the short-time slopes reproduce
$|p+q|$ from Eq.~(\ref{eq:slope0}) to within $0.1\%$ for all seven values
tested, while differing from $2\Lambda$ by factors up to $2.29$; this is the
most direct confirmation of the bias identified in Sec.~\ref{sec:gauss}.
Second, the asymptotic rates agree with Eq.~(\ref{eq:Lamofr}) to $2\%$ or
better at $\mathcal A\lesssim4$; the residual drift at large $\mathcal A$
reflects that the transient persists longer there, so that the window
$5\le T\le50$ is not yet fully asymptotic at $N=800$.

\begin{figure}
\includegraphics{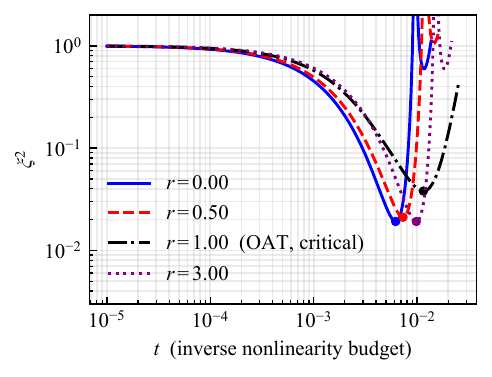}
\caption{\label{fig4}Squeezing versus time at $N=200$ for four Kerr ratios.
Every curve carries its own colour and line style: blue solid ($r=0$), red
dashed ($r=0.5$), black dash-dotted ($r=1$) and purple dotted ($r=3$).  The
black dash-dotted curve ($r=1$) is the critical case: there $\Lambda=0$ and
the decay is markedly slower and algebraic.  Dots mark the respective
optima.}
\end{figure}

Figure~\ref{fig4} follows the time evolution for four ratios at one system
size.  The critical case is clearly distinct from the others, its decay being
algebraic rather than exponential and its optimum arriving much later and
correspondingly shallower, whereas the three hyperbolic curves differ mainly
in how early their optima fall.

The optimal depths at $r=0$, $r=+3$ and
$r=-3$ coincide exactly, $\xi^2_{\min}=4.86514\times10^{-3}$ at $N=800$, and
the same holds at every other~$N$.  This key structural result is exact, and
its origin is a rotation.  At $r=0$ the Hamiltonian is proportional
to $\hat J_x^2-\hat J_y^2$ while the CSS points along~$z$; a rotation by
$\pi/2$ about~$y$ maps $(\hat J_x,\hat J_z)$ to $(-\hat J_z,\hat J_x)$ and
turns the Hamiltonian into one proportional to $\hat J_z^2-\hat J_y^2$ with
the CSS along~$x$, which is precisely the case $r=+3$, the case $r=-3$
following from the analogous rotation about~$x$.  The three points therefore
describe one and the same unitary orbit, and only the speed differs: at unit
budget $\Lambda/J=4.000$ at $r=0$ but $2.530$ at $r=\pm3$, because at
$r=\pm3$ a fraction $1/\sqrt{10}$ of the nonlinearity budget is spent on
self-Kerr.  Consequently $r=0$, that is, pure pair exchange, is the unique
point that is optimal in rate and in attainable depth at the same time.

\subsection{Scaling with $N$}

The exponent of the depth is the second quantity that distinguishes the two
classes of flow, so we extract it next.

\begin{figure}
\includegraphics[width=\columnwidth]{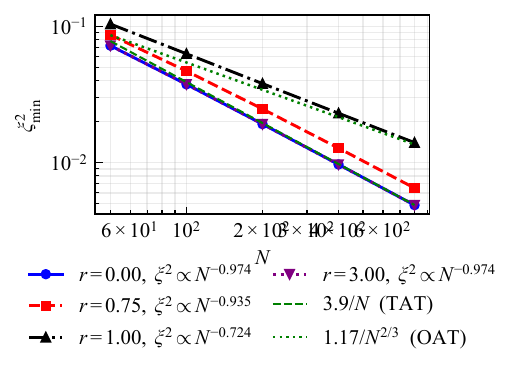}
\caption{\label{fig5}Optimal squeezing versus system size.  Markers joined by
lines: exact numerics, with the fitted exponents of Table~\ref{tab:depth}
quoted in the legend.  Green dashed and dotted lines are $3.9/N$ (TAT) and
$1.17/N^{2/3}$ (OAT).  The $r=3$ data (purple, dotted) lie exactly on top of
the $r=0$ data (blue, solid) and are distinguished from them by line style
alone.}
\end{figure}

Figure~\ref{fig5} and Table~\ref{tab:depth} give the
finite-size behavior.  Fitting $\xi^2_{\min}\propto N^{-p}$ over
$N=50$--$800$, we find $p=0.974$ at the balanced saddles $r\in\{0,\pm3\}$,
in excellent agreement with the TAT asymptote $p=1$, with a
numerical prefactor $3.89/N$ against the analytic $4/N$ of
Eq.~(\ref{eq:optTAT}), a difference consistent with finite-$J$ corrections.
At the critical points $r=\pm1$ the fit gives $p=0.724$, the expected OAT
value $2/3$ with finite-size corrections, whereas intermediate anisotropies
interpolate smoothly and give $p=0.87$--$0.97$, never exceeding $0.974$.

\subsection{Universality: depth depends on $\mathcal A$ alone}

The results so far are organized by the ratio~$r$, but they are in fact
organized by the anisotropy, as the following replotting shows.

\begin{figure*}
\includegraphics{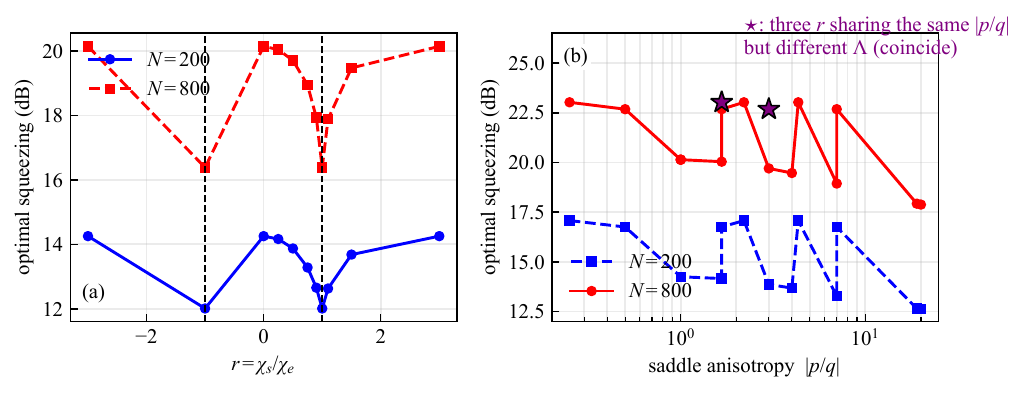}
\caption{\label{fig6}(a) Optimal squeezing versus Kerr ratio at two system
sizes.  The minima of the curve sit at the exact zeros of $\Lambda$,
$r=\pm1$, which the black dashed verticals and their
$\Lambda=0$ labels mark; the maxima sit at $r\in\{0,\pm3\}$.  Note that even
the worst point
is only $4.6$~dB below the best value at $N=800$.  (b) The same data replotted
against the saddle anisotropy $\mathcal A$, together with the additional
points of Table~\ref{tab:uni} (purple stars).  All data collapse onto single
monotone curves, demonstrating that the depth is a function of $\mathcal A$
alone and is independent of $\Lambda$.}
\end{figure*}

\begin{table}[b]
\caption{\label{tab:uni}Test of universality: groups of $r$ sharing
the same anisotropy $\mathcal A$ but having different rates $\Lambda$.
Within each group the attainable depth is identical,
although $\Lambda$ differs by up to $1.6\times$.}
\begin{ruledtabular}
\begin{tabular}{ c c d c c }
\multicolumn{1}{c}{$\mathcal A$}&
\multicolumn{1}{c}{$r$}&
\multicolumn{1}{c}{$\Lambda/J$}&
\multicolumn{1}{c}{depth ($N=200$)}&
\multicolumn{1}{c}{depth ($N=800$)}\\
\colrule
      & 0.5    & 3.0984 & 16.7574 & 22.6798\\
 3    & 5/3    & 2.3764 & 16.7574 & 22.6798\\
      & 7      & 1.9596 & 16.7574 & 22.6798\\
\colrule
      & 0.25   & 3.7573 & 17.0789 & 23.0264\\
 5/3  & 2.2    & 2.5642 & 17.0789 & 23.0264\\
      & 13/3   & 2.3223 & 17.0789 & 23.0264\\
\end{tabular}
\end{ruledtabular}
\end{table}

Figure~\ref{fig6}(a) shows the optimal depth versus~$r$: it is
largest at $r\in\{0,\pm3\}$ and smallest at the critical points $r=\pm1$,
rising monotonically between them with the distance $|r-1|$ until it
saturates.  The total variation is modest, the worst case at $N=800$ still
delivering $18.5$~dB against $23.1$~dB for the best, because the
dynamics at the critical points is one-axis twisting, which remains an
effective squeezer.  What is lost there is the rate: exponential growth is
replaced by algebraic growth, and the asymptotic exponent falls from $0.974$
to $0.724$.

Figure~\ref{fig6}(b) replots the same data against the anisotropy
$\mathcal A$ and reveals the organizing principle that, at fixed~$N$, the
depth is a single monotone function of~$\mathcal A$, for which
Table~\ref{tab:uni} gives the strongest evidence.
Equation~(\ref{eq:aniso}) implies that infinitely many
values of~$r$ share a given~$\mathcal A$: for instance, $\mathcal A=3$
occurs at $r=0.5$, $5/3$ and~$7$, while $\mathcal A=5/3$ occurs at
$r=0.25$, $2.2$ and~$13/3$.  These values of~$r$ have substantially different
rates, yet the exact numerical depths agree at both $N=200$ and $N=800$.  The
two quantities of Sec.~\ref{sec:mf} are therefore independent: $\Lambda$ sets
the clock and $\mathcal A$ sets the ceiling.

\subsection{\label{sec:bench}Benchmark against canonical one- and two-axis
twisting}

The identity $\hat E=2(\hat J_x^2-\hat J_y^2)$ means that the point $r=0$ is
two-axis twisting in its own right, reached by a different physical process,
and quantifying what this route gains over canonical one-axis
twisting requires a common scale.  We therefore compare the
generators $\hat J_+^2+\hat J_-^2$, $\hat J_x^2-\hat J_y^2$ and~$\hat J_z^2$
at equal Frobenius norm of the traceless part of the twisting tensor,
writing $\tilde\lambda_\mu=\lambda_\mu-\bar\lambda$ with
$\bar\lambda=\frac13(\lambda_x+\lambda_y+\lambda_z)$.  Adding multiples of
$\hat J^2$ does not matter, as noted after Eq.~(\ref{eq:Lambda}), and we set
$\|\tilde{\bm\lambda}\|=2\sqrt2$, the value realized by the
unit-budget $r=0$ point used throughout Sec.~\ref{sec:depth}, so that pair
exchange, canonical two-axis twisting and one-axis twisting all stand on the
same footing.

\begin{table}[b]
\caption{\label{tab:bmark}Controlled benchmark at equal nonlinearity budget,
$\|\tilde{\bm\lambda}\|=2\sqrt2$ (see text).  The $r=0$ family and canonical
two-axis twisting are one and the same time evolution; all differences from
one-axis twisting are therefore attributable to the twisting geometry
and not to interaction strength.}
\begin{ruledtabular}
\begin{tabular}{ c c c c c }
\multicolumn{1}{c}{generator}&
\multicolumn{1}{c}{$N$}&
\multicolumn{1}{c}{$\xi^2_{\min}$}&
\multicolumn{1}{c}{dB}&
\multicolumn{1}{c}{$t_{\rm opt}$}\\
\colrule
      & 100 & $3.75484\times10^{-2}$ & 14.254 & $1.067\times10^{-2}$\\
TAT   & 200 & $1.91597\times10^{-2}$ & 17.176 & $6.202\times10^{-3}$\\
($r=0$)& 400 & $9.67958\times10^{-3}$ & 20.141 & $3.536\times10^{-3}$\\
      & 800 & $4.86514\times10^{-3}$ & 23.129 & $1.985\times10^{-3}$\\
\colrule
      & 100 & $6.29463\times10^{-2}$ & 12.010 & $1.454\times10^{-2}$\\
OAT   & 200 & $3.80014\times10^{-2}$ & 14.202 & $9.385\times10^{-3}$\\
      & 400 & $2.30108\times10^{-2}$ & 16.381 & $6.021\times10^{-3}$\\
      & 800 & $1.40047\times10^{-2}$ & 18.537 & $3.845\times10^{-3}$\\
\end{tabular}
\end{ruledtabular}
\end{table}

\begin{figure*}
\includegraphics{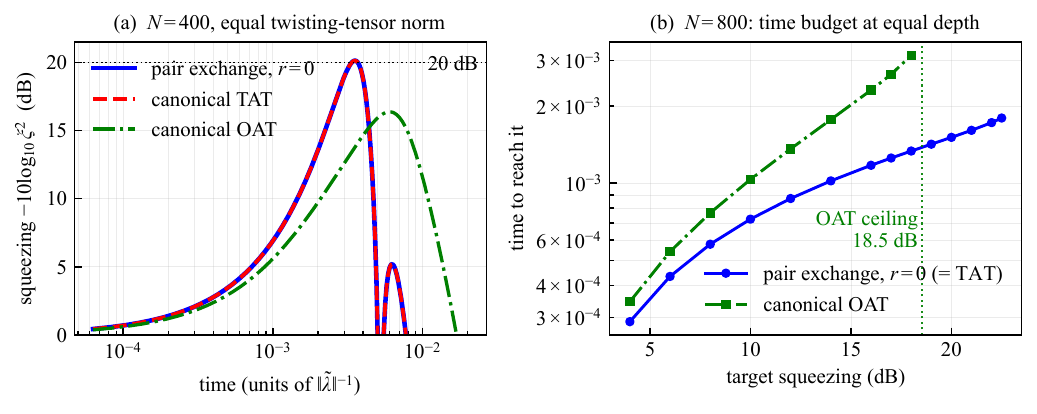}
\caption{\label{fig7}Benchmark against canonical twisting, all at the same
budget $\|\tilde{\bm\lambda}\|=2\sqrt2$.  (a) Squeezing versus time at
$N=400$.  The $r=0$ curve and the $\hat J_x^2-\hat J_y^2$ curve are the
same curve (they agree to machine precision at every sampled time); they are
drawn as the blue solid and red dashed lines that lie exactly on top of each
other, while one-axis twisting (green dash-dotted) falls away from them.  The
dotted horizontal guide is
$20$~dB.  (b) Time required to reach a target
depth at $N=800$.  Below about $12$~dB the two routes need comparable times,
but one-axis twisting saturates at its ceiling of $18.54$~dB, drawn as the
dotted vertical asymptote: the curve turns
vertical and no additional waiting time yields more squeezing.  Two-axis twisting reaches
$20$~dB in $1.51\times10^{-3}$ and continues to $23.13$~dB.}
\end{figure*}

Table~\ref{tab:bmark} and Fig.~\ref{fig7} summarize the comparison.  To
confirm that the two routes are one and the same time evolution rather than
merely similar, we sampled $\xi^2(t)$ at forty points spanning three decades
for $N=50$, $200$ and~$800$, finding no deviation beyond numerical
round-off between $\hat H=\chi_e(\hat J_+^2+\hat J_-^2)$ and
$\hat H=\chi(\hat J_x^2-\hat J_y^2)$ at matched budget.  The operator
identity itself holds to machine precision up to $N=200$.  Every number in
the upper block of Table~\ref{tab:bmark} is therefore at once a result about
pair exchange and a result about two-axis twisting.

The advantage over one-axis twisting is therefore an advantage in ceiling
first and in speed only second.  Fitting $\xi^2_{\min}=cN^{-p}$ over
$N=100$--$800$ gives $3.49\,N^{-0.983}$ for two-axis twisting against
$1.75\,N^{-0.723}$ for one-axis twisting, the prefactor depending on the
fitting window because the approach to the asymptote $p=1$ is logarithmic:
over $N=50$--$800$ one finds $3.89\,N^{-0.974}$, and at $N=800$ the two
differ by $4.59$~dB.  Figure~\ref{fig7}(b) makes the operative difference
plain: at a modest target the two routes cost comparable time, a factor
$1.43$ at $10$~dB and nearly independent of~$N$.  The OAT curve then
terminates, so that at $N=800$ one-axis twisting can never deliver $20$~dB
however long one waits, whereas two-axis twisting reaches $20$~dB in
$1.51\times10^{-3}$ inverse nonlinearity units and continues to $23.13$~dB.
The ratio of optimal times grows slowly,
$t_{\rm opt}^{\rm OAT}/t_{\rm opt}^{\rm TAT}=1.36,\,1.51,\,1.70$
and~$1.94$ at $N=100,\,200,\,400$ and~$800$, consistent with the $\ln J$
behavior of Eq.~(\ref{eq:optTAT}), while the ratio of attainable depths
diverges as $N^{0.26}$.  For a device with a fixed coherence time this is the
trade-off that matters: extra depth is available only from a Hamiltonian in
the hyperbolic class, whereas extra patience is not.

\subsection{\label{sec:wigner}Phase-space signature at the optimum}

Because the Wineland parameter reports only second moments, it is blind to
the difference between a genuinely Gaussian uncertainty ellipse and a
distribution that merely has the same covariance — a distinction that is not
academic, since it separates the regimes in which the closed-form solution of
Sec.~\ref{sec:gauss} is quantitative, $T\lesssim10$, from those in which it
is not.  We therefore examine the states directly on the generalized Bloch sphere
using the $SU(2)$-covariant Wigner function~\cite{Dowling1994,Brif1999},
\begin{equation}
W(\theta,\phi)=\sum_{k=0}^{2j}\sum_{q=-k}^{k}\rho_{kq}\,Y_{kq}(\theta,\phi),
\label{eq:Wdef}
\end{equation}
computed with the routines of Ref.~\onlinecite{Lambert2024}.  Here
$j\equiv J$ is the spin quantum number, $Y_{kq}(\theta,\phi)$ are the
spherical harmonics on the sphere, and $\rho_{kq}$ are the multipole moments
of the density operator, obtained as the expectation values of the
corresponding irreducible tensor operators.  We adopt units in which
$\int W\,d\Omega=1$, so that the maximally mixed state has $W=1/4\pi$.  A spin
coherent state occupies a patch whose angular variance, summed over the two
transverse directions, is $1/j$, and its peak lies at
$W\simeq(2j+1)/2\pi$, which evaluates to $4.93$ at $j=15$; the numerical
evaluation below returns $4.85$, the difference being a finite-$j$
correction.  As a scale-free measure of how far a state has moved away from
any classical distribution we use the negative-volume fraction
\begin{equation}
\delta=\frac{\int_{W<0}|W|\,d\Omega}{\int|W|\,d\Omega} ,
\label{eq:delta}
\end{equation}
which vanishes identically for any coherent state and grows monotonically with
twisting time.

Because comparing shapes requires comparing frames, each state is rotated
before evaluation so that (i)~$\langle\hat{\bm J}\rangle$ points along $+z$ and
(ii)~the minimum-variance direction lies along~$x$, both conditions being
verified numerically for every panel shown.  With this convention
the squeezed quadrature is horizontal and the anti-squeezed one vertical in
every panel, so that the four can be compared directly.

\begin{figure*}[tbh]
\includegraphics[width=\textwidth]{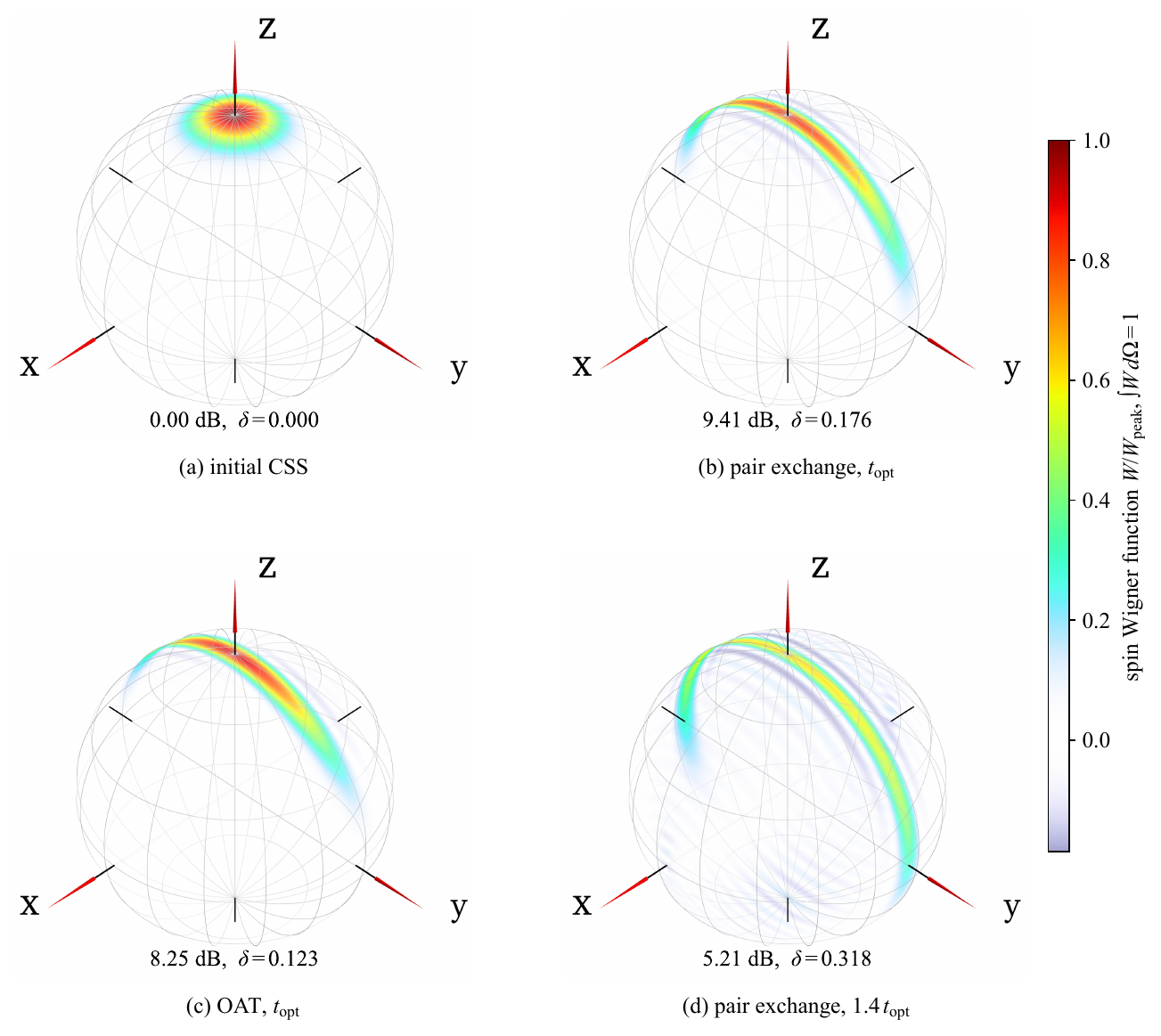}
\caption{\label{fig8}Spin Wigner function at $N=30$ ($j=15$), plotted on the
generalized Bloch sphere with the graticule and the Cartesian axes drawn for
orientation.  Each state is evaluated after rotation into the common frame
specified in the text, in which the mean spin points along $+z$ and the
squeezed quadrature along~$x$.  All four panels share a
single colour scale, normalized to the peak of the coherent state.  The
colour is therefore a direct readout of~$W$.  The sphere surface is white
wherever $W$ is essentially zero, and the bright patch around the $+z$ pole
carries the signal.  (a)~initial CSS: a circular patch of half-width
$12^{\circ}$, with zero negativity.  (b)~two-axis twisting realized by pair
exchange ($r=0$) at its optimum, $t_{\rm opt}=2.566\times10^{-2}$, giving
$9.41$~dB:
the patch is squeezed to $3.0^{\circ}$ along~$x$ and stretched to
$51^{\circ}$ along~$y$, and has begun to curve.  (c)~canonical
one-axis twisting at its optimum, $t_{\rm opt}=3.044\times10^{-2}$,
giving $8.25$~dB: $3.8^{\circ}\times40.5^{\circ}$.  (d)~two-axis
twisting at $1.4\,t_{\rm opt}$, where $\xi^2$ has fallen back to $5.21$~dB
while the anti-squeezed tail has wrapped $87^{\circ}$ around the sphere.  The
negative-volume fraction $\delta$, Eq.~(\ref{eq:delta}), is quoted in each
panel.}
\end{figure*}

Only the coherent state has vanishing negativity:
$\delta=0.0000$ in panel~(a), against $0.176$ in~(b), $0.123$ in~(c) and
$0.318$ in~(d).  This is the expected behavior.  Within the
$SU(2)$-covariant family the coherent states are the only positive
ones~\cite{Brif1999}.  The result is also a useful consistency check on the
whole computation, because $\delta$ uses the full~$W$ and not only its peak.
Panels~(b) and~(c) carry almost the same metrological gain, $9.41$~dB against
$8.25$~dB.  They pay for it differently.  Their mean-spin lengths differ,
$|\langle\hat{\bm J}\rangle|=0.764J$ against $0.851J$, and so do their aspect
ratios.  The reason is that $\xi^2$ is the product of two factors, how much the
variance shrinks and how much the mean spin shortens, and the two routes reach
a comparable figure of merit by spending differently on those two factors.
Panel~(d) is where the Gaussian description fails, and it fails in a way that
no variance-based measure signals: at $1.4\,t_{\rm opt}$ the squeezed
half-width is still $3.0^{\circ}$, four times better than the coherent-state
value, while the anti-squeezed tail has wrapped $87^{\circ}$
around the sphere and the mean spin has dropped to $0.446J$.  The Wineland
parameter correctly reports that squeezing is being lost, from $9.41$ to
$5.21$~dB, but it attributes the loss to contrast and does not display it as
curvature, whereas the phase-space picture shows that the state has simply
run out of sphere.  Geometrical saturation, not a change in the local
expansion rate, terminates the depth analysed in Sec.~\ref{sec:depth}, which
is why non-Gaussian states beyond the squeezing optimum have to be treated
with measures other than $\xi^2$~\cite{Strobel2014,Evrard2019,Baamara2021}.

Nothing in panels~(b)--(d) depends on how the state was
produced: since pair exchange at $r=0$ and canonical two-axis twisting agree
pointwise to machine precision, as Sec.~\ref{sec:bench} shows, this figure is
equally a statement about pair exchange and about two-axis twisting.

\section{\label{sec:disc}Discussion}

The preceding sections treated $r$ as a free parameter.  This section asks
how far it must be moved in practice, how the result relates to earlier work,
and which idealizations it rests on.

\subsection{How far the ratio has to be pushed from unity}

Appendices~\ref{app:N2} and~\ref{app:chi3} establish two facts, which
together explain why the constraint has received little attention.  First, at
$N=2$ the self-Kerr operator assigns identical diagonal energies to
$|2,0\rangle$ and $|0,2\rangle$, so that the pair-exchange transition is
perfectly resonant whatever the value of~$\chi_s$.  One-axis twisting and
two-axis twisting are then rigorously indistinguishable, and no experiment or
theory restricted to the $N\le2$ manifold can detect~$r$; indeed, every
proposal for two-photon exchange that we know of operates in a low-excitation
manifold and therefore never encounters the constraint.  The degeneracy is
first lifted at $N=4$, where the self-Kerr energy mismatch between
$|4,0\rangle$ and $|2,2\rangle$ is $8\chi_s$ against a pair-exchange matrix
element of
$\sqrt{(3\cdot4)(2\cdot1)}\,\chi_e=\sqrt{24}\,\chi_e=4.90\,\chi_e$, so that
once $\chi_s$ is comparable to~$\chi_e$ the whole ladder is driven out of
resonance.  That is the microscopic counterpart of $\Lambda(\pm1)=0$.

Second, an isotropic $\chi^{(3)}$ with two modes sharing a spatial profile
produces $\chi_s$ and $\chi_e$ in exactly the ratio $1:1$, so that the natural
operating point is the unfavourable one.  Table~\ref{tab:depth} and
Fig.~\ref{fig6} show what the requirement amounts to: a deviation from unity
rather than an elimination of self-Kerr.  A ratio of $0.9$ already restores
most of the deficit in the scaling exponent, which rises from $0.724$ to
$0.874$, and it reaches $20.7$~dB at $N=800$.  A ratio of $0.5$ comes within
$0.45$~dB of the optimal depth, $22.68$ against $23.13$~dB, and already shows
an exponent of $0.964$.  Nothing is gained by moving the ratio below zero,
since at $r=-3$ the depth equals the depth at $r=0$ but the rate is only
$0.63$ of it.  Equation~(\ref{eq:chis}) makes the experimental handle
explicit: because $\chi_s=U-\frac12U_c$, one has to shape the spatial overlap
integrals so that the cross-Kerr coefficient approaches twice the self-Kerr
coefficient, $U_c\to2U$.  Two modes with different transverse profiles
$u_a(\bm r)$ and $u_b(\bm r)$ in the same medium give independent control of
the four relevant overlaps, $\int|u_a|^4$, $\int|u_b|^4$, $\int|u_a|^2|u_b|^2$
and $\int u_a^{*2}u_b^2$.  This is therefore a mode-design problem rather than
a materials problem.

\subsection{Relation to prior work}

We now place the three results of Sec.~\ref{sec:intro} in the context of the
works listed there.  Of those, the one closest to our analysis
is that of Opatrn\'y~\cite{Opatrny2015}, who formulated twisting and twisting
rates for an arbitrary symmetric twisting tensor and, among other things,
obtained the two quantities that we use as items~(a) and~(b) of the
Introduction: the hyperbolic rate
$2[(\lambda_{\max}-\lambda_{\rm mid})(\lambda_{\rm
mid}-\lambda_{\min})]^{1/2}$ and the midpoint condition
$\lambda_{\rm mid}=\frac12(\lambda_{\max}+\lambda_{\min})$ that characterizes
TAT.
Our Eq.~(\ref{eq:Lambda}) is that rate evaluated on the constrained tensor of
a $\chi^{(3)}$ platform, and our Eq.~(\ref{eq:aniso}) is that condition turned
into a function of the single surviving ratio~$r$.  The step from a free
tensor to a one-parameter family is what converts a geometric classification
into a design rule, and it is also what makes the two thresholds fall at
different places.  With a free tensor one may sit at a balanced saddle and
have a good rate at the same time; here the only balanced points,
$r\in\{0,\pm3\}$, are not the points of best rate.  The independence of the
two conditions, item~(c) of the Introduction, becomes visible only after this
step, and we regard it as our main addition.

Hardal and M\"ustecapl{\i}o{\u g}lu~\cite{Hardal2014} studied a
two-site Kerr--Hubbard model with two-particle exchange, stating explicitly
that this coupling has an inter-cavity two-axis-twisting character.  They
noted that $2J(J_x^2-J_y^2)$ is the Lipkin--Meshkov--Glick
model~\cite{Lipkin1965}, solved the master equation for driven, dissipative
cavities and characterized the steady-state squeezing, entanglement and
coherence against the relative strength of the exchange and the nonlinearity.
Their Hamiltonian therefore contained exactly the two ingredients studied
here, and their scan over the relative strength is close in spirit to our scan
over~$r$.  What is absent from that analysis, and to our knowledge from the
literature generally, is any identification of two separate thresholds at
two different ratios, namely the exact vanishing of the rate at $r=\pm1$ and
the exact optimality of the depth at $r\in\{0,\pm3\}$, as well as the
closed-form solution Eqs.~(\ref{eq:Tdef})--(\ref{eq:xi2closed}) that explains
why the two are not the same number.  For a general tensor it is known that
the initial-stage rate and the asymptotic rate differ~\cite{Opatrny2015}; our
addition is that this difference reorganizes into two distinct sets of~$r$
once a $\chi^{(3)}$ platform constrains the tensor, and that the second of
those sets fixes a ceiling rather than a clock.  The distinction matters
because it changes the design target.  A scheme optimized for depth at $r=3$
would pass a naive test for two-axis twisting while directing about a third
of its nonlinearity budget to the self-Kerr term, which does not increase the
rate.

Our result also clarifies the status of several neighbouring activities.  The
extended Bose--Hubbard pair-tunnelling
literature~\cite{Folling2007,Zhu2015} studies the same operator, but aims at
ground-state phases and never at squeezing; where the same number-conserving
Hamiltonian is studied for metrology~\cite{Volkoff2016}, the pair-tunnelling
amplitude is the quantity to be estimated rather than a resource for
generating squeezing.  The work closest to our question
asks how fast and how optimally entanglement can be generated in a two-mode
bosonic system at fixed nonlinearity: in bosonic Josephson junctions
the optimal protocols and their speed limits have been mapped out in
detail~\cite{Micheli2003,Sorelli2019} and made robust against the readout by
undoing the twisting dynamics~\cite{Mirkhalaf2018}.  The aim differs from
ours, since there the two-mode structure converts a given nonlinearity into
entanglement as quickly as possible, whereas here we ask which kind of
nonlinearity the same platform presents to the collective spin in the first
place.  In circuit QED the engineered multi-photon
processes realized so far are of two types.  The first,
$g_2\hat a^2\hat b^\dagger+{\rm h.c.}$ with $g_2$ the coupling
strength~\cite{Leghtas2015}, converts two photons of one mode into one
photon of the other, a pump supplying the energy difference.  The second is
Raman-assisted hopping of a pair between the ladders of the two modes,
$|k,l\rangle\leftrightarrow|k+1,l+1\rangle$ with $k$ and~$l$ the occupations
of the two modes~\cite{Nguyen2024}; its bosonic form
$\hat a^\dagger\hat b^\dagger+\hat a\hat b$ is the two-mode squeezing
generator of $\mathrm{su}(1,1)$ rather than the $\mathrm{su}(2)$ twisting
generator studied here.  Reports of two-axis-twisting-like squeezing in such
systems therefore refer to the shape of the Husimi distribution rather than to
the generator.
Finally, the scarcity of pair exchange in practice has a simple origin: a
$\chi^{(3)}$ four-wave-mixing process involves four field amplitudes, and
promoting one of them to a classical pump leaves only three quantum fields,
so that the best one can obtain is a term of the form $\hat a^{\dagger2}\hat b$.
Producing $\hat a^{\dagger2}\hat b^2$ instead requires either two degenerate
modes and no pump, or a fifth-order process combined with a pump.  Correlated
two-photon sources obtained from cascaded emission do not evade this counting,
since they populate the $\mathrm{su}(1,1)$ pair-creation channel discussed
above rather than the number-conserving exchange required here.

\subsection{Limitations}

The conclusions above rest on several idealizations, which we state here.
Because our treatment is unitary and restricted to the fixed-$N$ sector,
photon loss, dephasing and pump depletion lie outside its scope.  Since the
advantage of TAT over OAT is largely one of time, TAT approaching the optimum
exponentially whereas OAT does so only algebraically, decoherence will
penalize the near-critical regime $r\approx\pm1$ more severely than our ideal
numbers suggest, so that the practical case for moving toward $r=0$ becomes
stronger rather than weaker.  We also note that Eq.~(\ref{eq:xi2closed}) is a
Gaussian, that is, linearized, result, which as Table~\ref{tab:model}
documents ceases to be quantitative once $T\sim J$; all statements about the
attainable depth therefore rest on exact diagonalization.  A further
limitation is that we have optimized $\xi^2$ throughout: past the optimum,
where the Gaussian picture breaks down, $\xi^2$ ceases to be the relevant
figure of merit, as the explicit comparison of the quantum Fisher information
and the squeezing parameter along the canonical two-axis-twisting evolution
shows~\cite{Kajtoch2015}, and deliberately non-Gaussian states are
known to outperform it both in the quantum Fisher information actually
attained~\cite{Strobel2014,Evrard2019,Zou2018} and in their
scaling~\cite{Baamara2021}.  Two questions remain open, namely whether the
phase-space differences documented in Sec.~\ref{sec:wigner} survive into that
regime, and whether the $r$-dependence of the depth map extends to it.
Mode-entangled ensembles additionally allow the gain realized in a single
ensemble to be redistributed among spatially separated ones~\cite{Malia2022},
a use of the state that our single-ensemble figure of merit does not capture.

\section{\label{sec:concl}Conclusion}

Pair exchange between two bosonic modes is an exact realization of two-axis
twisting, and it becomes useful only when the accompanying Kerr nonlinearity
is kept away from the exchange strength.  We have shown that the single ratio
$r=\chi_s/\chi_e$ governs two logically distinct properties of the dynamics:
the exponential rate~$\Lambda$, which vanishes exactly at
$r=\pm1$, and the attainable depth, which is a universal function of
the saddle anisotropy~$\mathcal A$ and is maximal only at
$r\in\{0,\pm3\}$, where $\mathcal A=1$.  The point $r=0$ is uniquely optimal
in both respects.  The closed-form solution of the transverse covariance
obtained along the way shows that the short-time decay rate is $|p+q|$ and
that this rate coincides with $2\Lambda$ only for balanced saddles, a caveat
that applies to any protocol that extracts twisting rates from early-time
data.  Since the natural isotropic $\chi^{(3)}$ sits exactly at the forbidden
value $r=1$ and the constraint is invisible at $N=2$, we expect these two
conditions to be useful criteria in the design of schemes that aim at
Heisenberg-limited spin squeezing from pair exchange.

\vphantom{x}

\appendix

\section{\label{app:N2}Indistinguishability of the two generators at two
quanta}

In the $N=2$ sector the basis may be taken as $|2,0\rangle$, $|1,1\rangle$
and $|0,2\rangle$.  From Eq.~(\ref{eq:idS}),
\begin{equation}
\hat S=\mathrm{diag}(2,0,2),
\end{equation}
whereas $\hat E$ connects only $|2,0\rangle\leftrightarrow|0,2\rangle$, with
matrix element $\langle 0,2|\hat E|2,0\rangle=\sqrt{(2\cdot1)(1\cdot2)}=\sqrt4=2$,
so that the two states that $\hat E$ couples receive identical
diagonal shifts $2\chi_s$ from the self-Kerr, which therefore acts as the
identity on the resonant pair.  The transition frequency therefore
does not depend on~$\chi_s$, and the population dynamics depends only on~$\chi_e$.  The
observable difference first appears at $N=4$, where the relevant chain
$|4,0\rangle\leftrightarrow|2,2\rangle\leftrightarrow|0,4\rangle$ has
self-Kerr energies
$S(|4,0\rangle)=S(|0,4\rangle)=12$ and $S(|2,2\rangle)=4$, so that the
energy mismatch is $8\chi_s$.  The coupling is
$\langle 4,0|\hat E|2,2\rangle=\sqrt{(3\cdot4)(2\cdot1)}=\sqrt{24}=4.90$, in
units of~$\chi_e$.  This is why a study restricted to the two-excitation
manifold could not have revealed the constraint discussed in this paper.

\section{\label{app:chi3}An isotropic $\chi^{(3)}$ gives exactly $r=1$}

Take two bosonic modes that are orthogonal polarizations of one spatial mode
in a medium whose third-order susceptibility tensor $\chi^{(3)}$ is isotropic.
Quantizing the standard nonlinear energy density, the interaction Hamiltonian
is proportional to $2(\hat{\bm{\mathcal E}}\cdot\hat{\bm{\mathcal E}}^*)^2+
(\hat{\bm{\mathcal E}}\cdot\hat{\bm{\mathcal E}})
(\hat{\bm{\mathcal E}}^*\cdot\hat{\bm{\mathcal E}}^*)$, where
$\hat{\bm{\mathcal E}}$ is the positive-frequency part of the electric-field
operator.  Write the field as
$\hat{\bm{\mathcal E}}\propto \bm e_x\hat a_x+\bm e_y\hat a_y$.  Here
$\hat a_x$ and $\hat a_y$ are the two modes denoted $\hat a$ and $\hat b$
above, and $\bm e_x,\bm e_y$ are orthonormal polarizations sharing the same
spatial profile.  The interaction Hamiltonian then becomes
\begin{equation}
\hat H_{\rm nl}\propto 2\hat N^2+(\hat a_x^2+\hat a_y^2)
(\hat a_x^{\dagger2}+\hat a_y^{\dagger2}) .
\label{eq:Hint2}
\end{equation}
Expanding the second term and using $\hat a^2\hat a^{\dagger2}
=(\hat n+1)(\hat n+2)$,
\begin{align}
(\hat a_x^2+\hat a_y^2)(\hat a_x^{\dagger2}+\hat a_y^{\dagger2})
&=\sum_{\mu}\bigl[\hat n_\mu(\hat n_\mu-1)+4\hat n_\mu+2\bigr]+\hat E
\nonumber\\
&=\hat S+4\hat N+4+\hat E .
\end{align}
Since $\hat N$ is conserved, the terms $2\hat N^2$, $4\hat N$ and $4$ are
constants, and Eq.~(\ref{eq:Hint2}) reduces to
\begin{equation}
\hat H_{\rm nl}\propto \hat S+\hat E + \text{const}.
\end{equation}
Comparing with Eq.~(\ref{eq:Hgen}) gives $U=1$, $U_c=0$ and $\chi_e=1$ in
the common proportionality units, although this reading is not unique: since
the combination $\hat S+2\hat n_a\hat n_b=\hat N(\hat N-1)$ is constant inside
the sector, any pair with $2U-U_c=2$ represents the same operator.  Either
way, Eq.~(\ref{eq:chis}) gives
\begin{equation}
\chi_s=U-\tfrac12U_c=1,\qquad r=\chi_s/\chi_e=1 ,
\end{equation}
exactly.  The same conclusion follows from Eqs.~(\ref{eq:ids}), which give
$\hat S+\hat E=\hat N^2-4\hat J_y^2$: writing
$\hat J_y=\frac12(\hat n_+-\hat n_-)$, where $\hat n_\pm=\hat
a_\pm^\dagger\hat a_\pm$ and $\hat a_\pm=(\hat a_x\mp i\hat a_y)/\sqrt2$ are the
circular modes, the Hamiltonian becomes $4\hat n_+\hat n_-$.  In the
circular basis there is no pair exchange at all, because angular momentum is
conserved, and what remains is one-axis twisting about the circular-polarization
axis.  This agrees with the known behavior of Kerr-based polarization
squeezing~\cite{GarciaFerrer2007}, which sits on the critical line and yields
algebraic rather than exponential squeezing.

\section{\label{app:num}Numerical details}

All operators are constructed directly in the Fock basis of the fixed-$N$
sector (dimension $N+1$), avoiding an error that arises if
$\hat a$ and $\hat b$ are built separately and then multiplied: neither
operator closes on the sector, and their truncated products misplace the
square-root factors.  For example, $(\hat a^\dagger)^2\hat b^2$ returns
$(N-k)$, whereas $\langle k+2|\hat E|k\rangle$ requires
$\sqrt{(N-k)(N-k-1)}$, and the discrepancy reaches $\sqrt2$ near the edge of the
sector.  In the number basis $|k\rangle\equiv|k,N-k\rangle$, where $k$
denotes the occupation of mode~$a$ and $N-k$ that of mode~$b$, the correctly
constructed matrix elements are
\begin{align}
\langle k|\hat J_+|k-1\rangle&=\sqrt{k(N-k+1)},\\
\langle k|\hat S|k\rangle&=k(k-1)+(N-k)(N-k-1),\\
\langle k+2|\hat E|k\rangle&=\sqrt{(k+1)(k+2)(N-k)(N-k-1)},
\end{align}
which satisfy the identities (\ref{eq:ids}) exactly.  The CSS along
the (possibly complex) eigenaxis~$k$ is the $+J$ eigenvector of~$\hat J_k$,
which we obtain by Hermitian diagonalization rather than real-symmetric
diagonalization, since taking the real part of $\hat J_y$ at this stage zeroes
the operator and must therefore be avoided.
The time evolution uses one diagonalization of $\hat H$ followed by dense
matrix--vector products, with all nine symmetrized second moments precomputed
in the energy basis, and squeezing is minimized over the plane transverse to
the instantaneous mean spin, as the Wineland parameter prescribes.  We obtain
the minimum-variance directions
by explicit Hermitian eigendecomposition of the $2\times2$ covariance in that
plane rather than from a hand-derived principal-angle formula, which with an
incorrect sign silently interchanges the squeezed and
anti-squeezed axes; we detected this only by cross-checking against the
widths of the Wigner distribution itself.

The spin Wigner functions of Sec.~\ref{sec:wigner} were computed with
QuTiP~\cite{Lambert2024,Johansson2012}.  Two details of the computation matter.
First, \texttt{qutip.spin\_wigner} returns a
function normalized so that $\int W\,d\Omega=\sqrt{4\pi/(2j+1)}$ rather than
the value stated in its own documentation; multiplying by
$\sqrt{(2j+1)/4\pi}$ gives the unit-normalized function used in
Fig.~\ref{fig8}, which we verified on every panel.
Second, each state is brought into the common frame by the two-step rotation
(\ref{eq:rot}) below, and we check both frame conditions numerically instead
of assuming them: the residual transverse component of
$\langle\hat{\bm J}\rangle$ vanishes, and so does the residual $y$-component
of the squeezed axis.
\begin{equation}
\hat U=\exp\!\bigl[-i\gamma\hat J_z\bigr]\,
\exp\!\bigl[-i\beta\,\hat{\bm n}\cdot\hat{\bm J}\bigr],
\label{eq:rot}
\end{equation}
where $\hat U$ is the unitary that carries the state into the common frame and
$\hat{\bm z}$ is the unit vector along~$z$.  The axis and angle of the first
rotation are $\hat{\bm n}\propto\langle\hat{\bm J}\rangle\times\hat{\bm z}$
and $\beta=\arctan(|\langle\hat{\bm J}\rangle\times\hat{\bm z}|,
\langle\hat J_z\rangle)$; the azimuthal angle $\gamma$ of the second rotation
is fixed subsequently by re-reading the squeezed axis after the first
rotation.  (These rotation angles are unrelated to the anisotropy parameter
$\alpha$ of Sec.~\ref{sec:gauss}.)

Figure~\ref{fig8} is rendered with Mayavi~\cite{Ramachandran2011}.  One
non-obvious setting matters for correctness: VTK's surface actor has lighting
enabled by default, which modulates every colour by the local illumination and
destroys the correspondence between colour and~$W$; we therefore set
\texttt{lighting=False} and kept the opacity ramp of the lookup table close to
unity, so that the printed colour is the colormap value of $W$ and not a
shaded version of it.

\begin{acknowledgments}
This work was supported by NSFC (Grant
No.~12205092), the Hunan Provincial Natural Science Foundation of China (Grant
Nos.~2026JJ50100) and the Scientific Research Fund of Hunan Provincial
Education Department (Grant No.~24B0670).
\end{acknowledgments}

\bibliography{refs}

@PREAMBLE{
 "\providecommand{\noopsort}[1]{}"
 # "\providecommand{\singleletter}[1]{#1}%"
}

@ARTICLE{Kitagawa1993,
   author       = "M. Kitagawa and M. Ueda",
   title        = "Squeezed spin states",
   journal      = "Phys.\ Rev.\ A",
   volume       = "47",
   pages        = "5138",
   year         = "1993",
   doi          = "10.1103/PhysRevA.47.5138",
}

@ARTICLE{Wineland1992,
   author       = "D. J. Wineland and J. J. Bollinger and W. M. Itano and
                   F. L. Moore and D. J. Heinzen",
   title        = "Spin squeezing and reduced quantum noise in spectroscopy",
   journal      = "Phys.\ Rev.\ A",
   volume       = "46",
   pages        = "R6797",
   year         = "1992",
   doi          = "10.1103/PhysRevA.46.R6797",
}

@ARTICLE{Ma2011,
   author       = "Jian Ma and Xiaoguang Wang and Chang-Pu Sun and Franco Nori",
   title        = "Quantum spin squeezing",
   journal      = "Phys.\ Rep.",
   volume       = "509",
   pages        = "89--165",
   year         = "2011",
   doi          = "10.1016/j.physrep.2011.08.003",
}

@ARTICLE{Pezze2018,
   author       = "Luca Pezz{\`e} and Augusto Smerzi and Markus K. Oberthaler
                   and Roman Schmied and Philipp Treutlein",
   title        = "Quantum metrology with nonclassical states of atomic ensembles",
   journal      = "Rev.\ Mod.\ Phys.",
   volume       = "90",
   pages        = "035005",
   year         = "2018",
   doi          = "10.1103/RevModPhys.90.035005",
}

@ARTICLE{Toth2014,
   author       = "G. T{\'o}th and I. Apellaniz",
   title        = "Quantum metrology from a quantum information science perspective",
   journal      = "J.\ Phys.\ A",
   volume       = "47",
   pages        = "424006",
   year         = "2014",
   doi          = "10.1088/1751-8113/47/42/424006",
}

@ARTICLE{Opatrny2015,
   author       = "Tom{\'a}{\v s} Opatrn{\'y}",
   title        = "Twisting tensor and spin squeezing",
   journal      = "Phys.\ Rev.\ A",
   volume       = "91",
   pages        = "053826",
   year         = "2015",
   doi          = "10.1103/PhysRevA.91.053826",
}

@ARTICLE{Kajtoch2016,
   author       = "D. Kajtoch and E. Witkowska",
   title        = "Spin squeezing in dipolar spinor condensates",
   journal      = "Phys.\ Rev.\ A",
   volume       = "93",
   pages        = "023627",
   year         = "2016",
   doi          = "10.1103/PhysRevA.93.023627",
}

@ARTICLE{Hardal2014,
   author       = "Ali {\"U}.\ C. Hardal and {\"O}zg{\"u}r E. M{\"u}stecapl{\i}o{\u g}lu",
   title        = "Spin squeezing, entanglement, and coherence in two driven,
                   dissipative, nonlinear cavities coupled with single- and
                   two-photon exchange",
   journal      = "J.\ Opt.\ Soc.\ Am.\ B",
   volume       = "31",
   pages        = "1402--1409",
   year         = "2014",
   eprint       = "1303.4318",
   doi          = "10.1364/JOSAB.31.001402",
}

@ARTICLE{Alexanian2011,
   author       = "Moorad Alexanian",
   title        = "Two-photon exchange between two three-level atoms in separate cavities",
   journal      = "Phys.\ Rev.\ A",
   volume       = "83",
   pages        = "023814",
   year         = "2011",
   doi          = "10.1103/PhysRevA.83.023814",
}

@ARTICLE{Alexanian2011b,
   author       = "Moorad Alexanian",
   title        = "Dynamical generation of maximally entangled states in two
                   identical cavities",
   journal      = "Phys.\ Rev.\ A",
   volume       = "84",
   pages        = "052302",
   year         = "2011",
   doi          = "10.1103/PhysRevA.84.052302",
}

@ARTICLE{Muessel2015,
   author       = "W. Muessel and H. Strobel and D. Linnemann and
                   T. Zibold and B. Juli{\'a}-D{\'i}az and
                   M. K. Oberthaler",
   title        = "Twist-and-turn spin squeezing in {B}ose--{E}instein
                   condensates",
   journal      = "Phys.\ Rev.\ A",
   volume       = "92",
   pages        = "023603",
   year         = "2015",
   eprint       = "1507.02930",
   doi          = "10.1103/PhysRevA.92.023603",
}

@ARTICLE{Liu2011,
   author       = "Y. C. Liu and Z. F. Xu and G. R. Jin and L. You",
   title        = "Spin squeezing: Transforming one-axis twisting into
                   two-axis twisting",
   journal      = "Phys.\ Rev.\ Lett.",
   volume       = "107",
   pages        = "013601",
   year         = "2011",
   doi          = "10.1103/PhysRevLett.107.013601",
}

@ARTICLE{Luo2025,
   author       = "Chengyi Luo and Haoqing Zhang and Anjun Chu and
                   Chitose Maruko and Ana Maria Rey and James K. Thompson",
   title        = "Hamiltonian engineering of collective {XYZ} spin models in
                   an optical cavity",
   journal      = "Nat.\ Phys.",
   volume       = "21",
   pages        = "916--923",
   year         = "2025",
   doi          = "10.1038/s41567-025-02866-0",
}

@ARTICLE{Folling2007,
   author       = "S. F{\"o}lling and S. Trotzky and P. Cheinet and M. Feld
                   and R. Saers and A. Widera and T. M{\"u}ller and I. Bloch",
   title        = "Direct observation of second-order atom tunnelling",
   journal      = "Nature",
   volume       = "448",
   pages        = "1029--1032",
   year         = "2007",
   doi          = "10.1038/nature06112",
}

@ARTICLE{Zhu2015,
   author       = "Qizhong Zhu and Qi Zhang and Biao Wu",
   title        = "Extended two-site {B}ose--{H}ubbard model with pair
                   tunneling: spontaneous symmetry breaking, effective
                   ground state and fragmentation",
   journal      = "J.\ Phys.\ B",
   volume       = "48",
   pages        = "045301",
   year         = "2015",
   doi          = "10.1088/0953-4075/48/4/045301",
}

@ARTICLE{Leghtas2015,
   author       = "Zaki Leghtas and Steven Touzard and Ioan M. Pop and
                   Angela Kou and Brian Vlastakis and Andrei Petrenko and
                   Katrina M. Sliwa and Anirudh Narla and Shyam Shankar and
                   Michael J. Hatridge and Matthew Reagor and Luigi Frunzio
                   and Robert J. Schoelkopf and Mazyar Mirrahimi and
                   Michel H. Devoret",
   title        = "Confining the state of light to a quantum manifold by
                   engineered two-photon loss",
   journal      = "Science",
   volume       = "347",
   pages        = "853--857",
   year         = "2015",
   doi          = "10.1126/science.aaa2085",
}

@ARTICLE{Nguyen2024,
   author       = "Long B. Nguyen and Noah Goss and Karthik Siva and Yosep Kim
                   and Ed Younis and Bingcheng Qing and Akel Hashim and
                   David I. Santiago and Irfan Siddiqi",
   title        = "Empowering a qudit-based quantum processor by traversing
                   the dual bosonic ladder",
   journal      = "Nat.\ Commun.",
   volume       = "15",
   pages        = "7117",
   year         = "2024",
   doi          = "10.1038/s41467-024-51434-2",
}

@ARTICLE{Lipkin1965,
   author       = "H. J. Lipkin and N. Meshkov and A. J. Glick",
   title        = "Validity of many-body approximation methods for a solvable
                   model. {I}. {E}xact solutions and perturbation theory",
   journal      = "Nucl.\ Phys.",
   volume       = "62",
   pages        = "188--198",
   year         = "1965",
   doi          = "10.1016/0029-5582(65)90862-X",
}

@ARTICLE{Schwinger1952,
   author       = "J. Schwinger",
   title        = "On angular momentum",
   journal      = "USAEC Report No.\ NYO-3071",
   year         = "1952",
   note         = "reprinted in L. C. Biedenharn and H. van Dam,
                   \textit{Quantum Theory of Angular Momentum}
                   (Academic Press, New York, 1965)",
}

@ARTICLE{Kitagawa1986,
   author       = "M. Kitagawa and Y. Yamamoto",
   title        = "Number-phase minimum-uncertainty state with reduced
                   number uncertainty in a {K}err nonlinear interferometer",
   journal      = "Phys.\ Rev.\ A",
   volume       = "34",
   pages        = "3974",
   year         = "1986",
   doi          = "10.1103/PhysRevA.34.3974",
}

@ARTICLE{Muessel2014,
   author       = "W. Muessel and H. Strobel and D. Linnemann and
                   D. B. Hume and M. K. Oberthaler",
   title        = "Scalable spin squeezing for quantum-enhanced
                   magnetometry with {B}ose--{E}instein condensates",
   journal      = "Phys.\ Rev.\ Lett.",
   volume       = "113",
   pages        = "103004",
   year         = "2014",
   doi          = "10.1103/PhysRevLett.113.103004",
}

@ARTICLE{Strobel2014,
   author       = "H. Strobel and W. Muessel and D. Linnemann and
                   T. Zibold and D. B. Hume and L. Pezz{\`e} and
                   A. Smerzi and M. K. Oberthaler",
   title        = "{F}isher information and entanglement of non-{G}aussian
                   spin states",
   journal      = "Science",
   volume       = "345",
   pages        = "424--427",
   year         = "2014",
   doi          = "10.1126/science.1250147",
}

@ARTICLE{Esteve2008,
   author       = "J. Est{\`e}ve and C. Gross and A. Weller and
                   S. Giovanazzi and M. K. Oberthaler",
   title        = "Squeezing and entanglement in a {B}ose--{E}instein
                   condensate",
   journal      = "Nature",
   volume       = "455",
   pages        = "1216--1219",
   year         = "2008",
   doi          = "10.1038/nature07332",
}

@ARTICLE{Gross2010,
   author       = "C. Gross and T. Zibold and E. Nicklas and J. Est{\`e}ve
                   and M. K. Oberthaler",
   title        = "Nonlinear atom interferometer surpasses classical
                   precision limit",
   journal      = "Nature",
   volume       = "464",
   pages        = "1165--1169",
   year         = "2010",
   doi          = "10.1038/nature08919",
}

@ARTICLE{Riedel2010,
   author       = "M. F. Riedel and P. B{\"o}hi and Y. Li and
                   T. W. H{\"a}nsch and A. Sinatra and P. Treutlein",
   title        = "Atom-chip-based generation of entanglement for quantum
                   metrology",
   journal      = "Nature",
   volume       = "464",
   pages        = "1170--1173",
   year         = "2010",
   doi          = "10.1038/nature08988",
}

@ARTICLE{Lucke2011,
   author       = "B. L{\"u}cke and M. Scherer and J. Kruse and
                   L. Pezz{\`e} and F. Deuretzbacher and P. Hyllus and
                   O. Topic and J. Peise and W. Ertmer and J. Arlt and
                   L. Santos and A. Smerzi and C. Klempt",
   title        = "Twin matter waves for interferometry beyond the
                   classical limit",
   journal      = "Science",
   volume       = "334",
   pages        = "773--776",
   year         = "2011",
   doi          = "10.1126/science.1208798",
}

@ARTICLE{Hamley2012,
   author       = "C. D. Hamley and C. S. Gerving and T. M. Hoang and
                   E. M. Bookjans and M. S. Chapman",
   title        = "Spin-nematic squeezed vacuum in a quantum gas",
   journal      = "Nat.\ Phys.",
   volume       = "8",
   pages        = "305--308",
   year         = "2012",
   doi          = "10.1038/nphys2245",
}

@ARTICLE{Sewell2012,
   author       = "R. J. Sewell and M. Koschorreck and M. Napolitano and
                   B. Dubost and N. Behbood and M. W. Mitchell",
   title        = "Magnetic sensitivity beyond the projection noise limit
                   by spin squeezing",
   journal      = "Phys.\ Rev.\ Lett.",
   volume       = "109",
   pages        = "253605",
   year         = "2012",
   doi          = "10.1103/PhysRevLett.109.253605",
}

@ARTICLE{Linnemann2016,
   author       = "D. Linnemann and H. Strobel and W. Muessel and
                   J. Schulz and R. J. Lewis-Swan and K. V. Kheruntsyan
                   and M. K. Oberthaler",
   title        = "Quantum-enhanced sensing based on time reversal of
                   nonlinear dynamics",
   journal      = "Phys.\ Rev.\ Lett.",
   volume       = "117",
   pages        = "013001",
   year         = "2016",
   doi          = "10.1103/PhysRevLett.117.013001",
}

@ARTICLE{Kruse2016,
   author       = "I. Kruse and K. Lange and J. Peise and B. L{\"u}cke and
                   L. Pezz{\`e} and J. Arlt and W. Ertmer and C. Lisdat and
                   L. Santos and A. Smerzi and C. Klempt",
   title        = "Improvement of an atomic clock using squeezed vacuum",
   journal      = "Phys.\ Rev.\ Lett.",
   volume       = "117",
   pages        = "143004",
   year         = "2016",
   doi          = "10.1103/PhysRevLett.117.143004",
}

@ARTICLE{Cox2016,
   author       = "K. C. Cox and G. P. Greve and J. M. Weiner and
                   J. K. Thompson",
   title        = "Deterministic squeezed states with collective
                   measurements and cavity feedback",
   journal      = "Phys.\ Rev.\ Lett.",
   volume       = "116",
   pages        = "093602",
   year         = "2016",
   doi          = "10.1103/PhysRevLett.116.093602",
}

@ARTICLE{Hosten2016,
   author       = "O. Hosten and R. Krishnakumar and N. J. Engelsen and
                   M. A. Kasevich",
   title        = "Quantum phase magnification",
   journal      = "Science",
   volume       = "352",
   pages        = "1552--1555",
   year         = "2016",
   doi          = "10.1126/science.aaf3397",
}

@ARTICLE{Evrard2019,
   author       = "A. Evrard and V. Makhalov and T. Chalopin and
                   L. A. Sidorenkov and J. Dalibard and R. Lopes and
                   S. Nascimbene",
   title        = "Enhanced magnetic sensitivity with non-{G}aussian spin
                   states",
   journal      = "Phys.\ Rev.\ Lett.",
   volume       = "122",
   pages        = "173601",
   year         = "2019",
   doi          = "10.1103/PhysRevLett.122.173601",
}

@ARTICLE{Zou2018,
   author       = "Yi-Quan Zou and Ling-Na Wu and Qi Liu and Xin-Yu Luo and
                   Shuai-Feng Guo and Jia-Hao Cao and Meng Khoon Tey and
                   Li You",
   title        = "Beating the classical precision limit with spin-1
                   {D}icke states of more than 10,000 atoms",
   journal      = "Proc.\ Natl.\ Acad.\ Sci.\ U.S.A.",
   volume       = "115",
   pages        = "6381--6385",
   year         = "2018",
   doi          = "10.1073/pnas.1715105115",
}

@ARTICLE{Mirkhalaf2018,
   author       = "Safoura S. Mirkhalaf and Samuel P. Nolan and
                   Simon A. Haine",
   title        = "Robustifying twist-and-turn entanglement with
                   interaction-based readout",
   journal      = "Phys.\ Rev.\ A",
   volume       = "97",
   pages        = "053618",
   year         = "2018",
   doi          = "10.1103/PhysRevA.97.053618",
}

@ARTICLE{Micheli2003,
   author       = "A. Micheli and D. Jaksch and J. I. Cirac and
                   P. Zoller",
   title        = "Many-particle entanglement in two-component
                   {B}ose--{E}instein condensates",
   journal      = "Phys.\ Rev.\ A",
   volume       = "67",
   pages        = "013607",
   year         = "2003",
   doi          = "10.1103/PhysRevA.67.013607",
}

@ARTICLE{Sorelli2019,
   author       = "G. Sorelli and M. Gessner and A. Smerzi and
                   L. Pezz{\`e}",
   title        = "Fast and optimal generation of entanglement in bosonic
                   {J}osephson junctions",
   journal      = "Phys.\ Rev.\ A",
   volume       = "99",
   pages        = "022329",
   year         = "2019",
   doi          = "10.1103/PhysRevA.99.022329",
}

@ARTICLE{Leroux2010,
   author       = "I. D. Leroux and M. H. Schleier-Smith and
                   V. Vuleti{\'c}",
   title        = "Implementation of cavity squeezing of a collective
                   atomic spin",
   journal      = "Phys.\ Rev.\ Lett.",
   volume       = "104",
   pages        = "073602",
   year         = "2010",
   doi          = "10.1103/PhysRevLett.104.073602",
}

@ARTICLE{SchleierSmith2010,
   author       = "M. H. Schleier-Smith and I. D. Leroux and
                   V. Vuleti{\'c}",
   title        = "Squeezing the collective spin of a dilute atomic
                   ensemble by cavity feedback",
   journal      = "Phys.\ Rev.\ A",
   volume       = "81",
   pages        = "021804",
   year         = "2010",
   doi          = "10.1103/PhysRevA.81.021804",
}

@ARTICLE{Pedrozo2020,
   author       = "E. Pedrozo-Pe{\~n}afiel and S. Colombo and C. Shu and
                   A. F. Adiyatullin and Z. Li and E. Mendez and
                   B. Braverman and A. Kawasaki and D. Akamatsu and
                   Y. Xiao and V. Vuleti{\'c}",
   title        = "Entanglement on an optical atomic-clock transition",
   journal      = "Nature",
   volume       = "588",
   pages        = "414--418",
   year         = "2020",
   doi          = "10.1038/s41586-020-3006-1",
}

@ARTICLE{Kirchmair2013,
   author       = "G. Kirchmair and B. Vlastakis and Z. Leghtas and
                   S. E. Nigg and H. Paik and E. Ginossar and
                   M. Mirrahimi and L. Frunzio and S. M. Girvin and
                   R. J. Schoelkopf",
   title        = "Observation of quantum state collapse and revival due
                   to the single-photon {K}err effect",
   journal      = "Nature",
   volume       = "495",
   pages        = "205--209",
   year         = "2013",
   doi          = "10.1038/nature11902",
}

@ARTICLE{Sivak2019,
   author       = "V. Sivak and N. Frattini and V. Joshi and
                   A. Lingenfelter and S. Shankar and M. Devoret",
   title        = "{K}err-free three-wave mixing in superconducting
                   quantum circuits",
   journal      = "Phys.\ Rev.\ Appl.",
   volume       = "11",
   pages        = "054060",
   year         = "2019",
   doi          = "10.1103/PhysRevApplied.11.054060",
}

@ARTICLE{Frattini2018,
   author       = "N. E. Frattini and V. V. Sivak and A. Lingenfelter and
                   S. Shankar and M. H. Devoret",
   title        = "Optimizing the nonlinearity and dissipation of a
                   {SNAIL} parametric amplifier for dynamic range",
   journal      = "Phys.\ Rev.\ Appl.",
   volume       = "10",
   pages        = "054020",
   year         = "2018",
   doi          = "10.1103/PhysRevApplied.10.054020",
}

@ARTICLE{Grimm2020,
   author       = "A. Grimm and N. E. Frattini and S. Puri and
                   S. O. Mundhada and S. Touzard and M. Mirrahimi and
                   S. M. Girvin and S. Shankar and M. H. Devoret",
   title        = "Stabilization and operation of a {K}err-cat qubit",
   journal      = "Nature",
   volume       = "584",
   pages        = "205--209",
   year         = "2020",
   doi          = "10.1038/s41586-020-2587-z",
}

@ARTICLE{Lescanne2020,
   author       = "Rapha{\"e}l Lescanne and Marius Villiers and
                   Th{\'e}au Peronnin and Alain Sarlette and
                   Matthieu Delbecq and Benjamin Huard and Takis Kontos and
                   Mazyar Mirrahimi and Zaki Leghtas",
   title        = "Exponential suppression of bit-flips in a qubit encoded
                   in an oscillator",
   journal      = "Nat.\ Phys.",
   volume       = "16",
   pages        = "509--513",
   year         = "2020",
   doi          = "10.1038/s41567-020-0824-x",
}

@ARTICLE{Dowling1994,
   author       = "J. P. Dowling and G. S. Agarwal and W. P. Schleich",
   title        = "{W}igner distribution of a general angular-momentum
                   state: {A}pplications to a collection of two-level
                   atoms",
   journal      = "Phys.\ Rev.\ A",
   volume       = "49",
   pages        = "4101--4109",
   year         = "1994",
   doi          = "10.1103/PhysRevA.49.4101",
}

@ARTICLE{Brif1999,
   author       = "C. Brif and A. Mann",
   title        = "Phase-space formulation of quantum mechanics and
                   quantum-state reconstruction for physical systems with
                   {L}ie-group symmetries",
   journal      = "Phys.\ Rev.\ A",
   volume       = "59",
   pages        = "971--987",
   year         = "1999",
   doi          = "10.1103/PhysRevA.59.971",
}

@ARTICLE{Davis2021,
   author       = "Jack Davis and Meenu Kumari and Robert B. Mann and
                   Shohini Ghose",
   title        = "{W}igner negativity in spin-$j$ systems",
   journal      = "Phys.\ Rev.\ Res.",
   volume       = "3",
   pages        = "033134",
   year         = "2021",
   doi          = "10.1103/PhysRevResearch.3.033134",
}

@ARTICLE{Troullinou2021,
   author       = "C. Troullinou and R. Jim{\'e}nez-Mart{\'i}nez and
                   J. Kong and V. G. Lucivero and M. W. Mitchell",
   title        = "Squeezed-light enhancement and backaction evasion in a
                   high sensitivity optically pumped magnetometer",
   journal      = "Phys.\ Rev.\ Lett.",
   volume       = "127",
   pages        = "193601",
   year         = "2021",
   doi          = "10.1103/PhysRevLett.127.193601",
}

@ARTICLE{Baamara2021,
   author       = "Y. Baamara and A. Sinatra and M. Gessner",
   title        = "Scaling laws for the sensitivity enhancement of
                   non-{G}aussian spin states",
   journal      = "Phys.\ Rev.\ Lett.",
   volume       = "127",
   pages        = "160501",
   year         = "2021",
   doi          = "10.1103/PhysRevLett.127.160501",
}

@ARTICLE{Huang2023,
   author       = "M.-Z. Huang and J. A. de la Paz and T. Mazzoni and
                   K. Ott and P. Rosenbusch and A. Sinatra and
                   C. L. Garrido Alzar and J. Reichel",
   title        = "Observing spin-squeezed states under spin-exchange
                   collisions for a second",
   journal      = "{PRX} {Q}uantum",
   volume       = "4",
   pages        = "020322",
   year         = "2023",
   doi          = "10.1103/PRXQuantum.4.020322",
}

@ARTICLE{Mao2023,
   author       = "T.-W. Mao and Q. Liu and X.-W. Li and J.-H. Cao and
                   F. Chen and W.-X. Xu and M. K. Tey and Y.-X. Huang and
                   L. You",
   title        = "Quantum-enhanced sensing by echoing spin-nematic
                   squeezing in atomic {B}ose--{E}instein condensate",
   journal      = "Nat.\ Phys.",
   volume       = "19",
   pages        = "1585--1590",
   year         = "2023",
   doi          = "10.1038/s41567-023-02168-3",
}

@ARTICLE{Malia2022,
   author       = "B. K. Malia and Y. Wu and J. Mart{\'i}nez-Rinc{\'o}n and
                   M. A. Kasevich",
   title        = "Distributed quantum sensing with mode-entangled
                   spin-squeezed atomic states",
   journal      = "Nature",
   volume       = "612",
   pages        = "661--665",
   year         = "2022",
   doi          = "10.1038/s41586-022-05363-z",
}

@ARTICLE{Greve2022,
   author       = "G. P. Greve and C. Luo and B. Wu and J. K. Thompson",
   title        = "Entanglement-enhanced matter-wave interferometry in a
                   high-finesse cavity",
   journal      = "Nature",
   volume       = "610",
   pages        = "472--477",
   year         = "2022",
   doi          = "10.1038/s41586-022-05197-9",
}

@ARTICLE{Robinson2024,
   author       = "J. M. Robinson and M. Miklos and Y. M. Tso and
                   C. J. Kennedy and T. Bothwell and D. Kedar and
                   J. K. Thompson and J. Ye",
   title        = "Direct comparison of two spin-squeezed optical clock
                   ensembles at the $10^{-17}$ level",
   journal      = "Nat.\ Phys.",
   volume       = "20",
   pages        = "208--213",
   year         = "2024",
   doi          = "10.1038/s41567-023-02310-1",
}

@ARTICLE{Lambert2024,
   author       = "N. Lambert and E. Gigu{\`e}re and P. Menczel and
                   B. Li and P. Hopf and G. Su{\'a}rez and M. Gali and
                   J. Lishman and R. Gadhvi and R. Agarwal and A. Galicia
                   and N. Shammah and P. Nation and J. R. Johansson and
                   S. Ahmed and S. Cross and A. Pitchford and F. Nori",
   title        = "{Q}u{T}i{P} 5: {T}he {Q}uantum {T}oolbox in {P}ython",
   journal      = "Phys.\ Rep.",
   volume       = "1153",
   pages        = "1--62",
   year         = "2026",
   eprint       = "2412.04705",
   doi          = "10.1016/j.physrep.2025.10.001",
}

@ARTICLE{Ramachandran2011,
   author       = "Prabhu Ramachandran and Gael Varoquaux",
   title        = "{M}ayavi: 3{D} visualization of scientific data",
   journal      = "Comput.\ Sci.\ Eng.",
   volume       = "13",
   pages        = "40--51",
   year         = "2011",
   doi          = "10.1109/MCSE.2011.35",
}

@ARTICLE{Johansson2012,
   author       = "J. R. Johansson and P. D. Nation and Franco Nori",
   title        = "{Q}u{T}i{P}: {A}n open-source {P}ython framework for the
                   dynamics of open quantum systems",
   journal      = "Comput.\ Phys.\ Commun.",
   volume       = "183",
   pages        = "1760--1772",
   year         = "2012",
   doi          = "10.1016/j.cpc.2012.02.021",
}

@ARTICLE{Qiao2026,
   author       = "Feng Qiao and Zu-Jian Ying",
   title        = "Phonon-induced two-axis spin squeezing with decoherence
                   reduction in a hybrid spin-optomechanical system",
   journal      = "Phys.\ Rev.\ A",
   volume       = "113",
   pages        = "043503",
   year         = "2026",
   doi          = "10.1103/cmb2-sktt",
   eprint       = "2510.07068",
   archivePrefix = "arXiv",
}

@ARTICLE{Ho2026,
   author       = "Le Bin Ho and Vu Xuan Tung Duong and Nozomu Takahashi and
                   Hiroaki Matsueda",
   title        = "Interference-induced state engineering and Hamiltonian
                   control for noisy collective-spin metrology",
   journal      = "Phys.\ Rev.\ A",
   volume       = "114",
   pages        = "032206",
   year         = "2026",
   doi          = "10.1103/sxxg-77b1",
   eprint       = "2603.22734",
   archivePrefix = "arXiv",
}

@ARTICLE{Koenig2026,
   author       = "J. D. Koenig and G. Barbieri and F. Fani Sani and C. A. Potts
                   and M. Kounalakis and G. A. Steele",
   title        = "Flux-modulated tunable interaction regimes in two strongly
                   nonlinear oscillators",
   journal      = "Phys.\ Rev.\ Research",
   volume       = "8",
   year         = "2026",
   doi          = "10.1103/3pf2-8n9b",
}

@ARTICLE{GarciaFerrer2007,
   author       = "F. V. Garcia-Ferrer and I. P{\'e}rez-Arjona and
                   G. J. de Valc{\'a}rcel and E. Rold{\'a}n",
   title        = "Quadrature and polarization squeezing in a dispersive optical
                   bistability model",
   journal      = "Phys.\ Rev.\ A",
   volume       = "75",
   pages        = "063823",
   year         = "2007",
   doi          = "10.1103/PhysRevA.75.063823",
}

@ARTICLE{Volkoff2016,
   author       = "T. J. Volkoff",
   title        = "Optimal and near-optimal probe states for quantum metrology of
                   number-conserving two-mode bosonic Hamiltonians",
   journal      = "Phys.\ Rev.\ A",
   volume       = "94",
   pages        = "042327",
   year         = "2016",
   doi          = "10.1103/PhysRevA.94.042327",
}

@ARTICLE{Kajtoch2015,
   author       = "D. Kajtoch and E. Witkowska",
   title        = "Quantum dynamics generated by the two-axis counter-twisting
                   Hamiltonian",
   journal      = "Phys.\ Rev.\ A",
   volume       = "92",
   pages        = "013623",
   year         = "2015",
   doi          = "10.1103/PhysRevA.92.013623",
}

\end{document}